\documentclass[aps,prd,amsmath,amssymb,preprintnumbers,nofootinbib,preprint,superscriptaddress]{revtex4-1}
\pdfoutput=1
\usepackage[utf8]{inputenc} %utf8
\usepackage{graphicx}
\graphicspath{{./figures/}}
\usepackage{url}
\usepackage[bookmarks, pagebackref=false]{hyperref}
\usepackage[usenames,dvipsnames]{xcolor}
\definecolor{orange}{cmyk}{0,0.5,1,0}
\definecolor{rossoCP3}{cmyk}{0,.88,.77,.40}
\definecolor{graa}{rgb}{0.8,0.8,0.8}
\definecolor{blaa}{rgb}{0.2,0.2,0.6}
\hypersetup{
	colorlinks, 
	bookmarksopen, 
	bookmarksnumbered,
	citecolor=blaa, 		%color of links to bibliography
	linkcolor=rossoCP3,	%color of internal links
	urlcolor=rossoCP3,			%color of external links 
	    }

\usepackage{amsthm}
\usepackage{bm}% bold math
\usepackage{bbm}
\usepackage{pxfonts}

\usepackage{amsmath,amssymb,amsfonts}
\usepackage{color}
\usepackage{float}
\usepackage{hyperref}
\usepackage[Symbolsmallscale]{upgreek}
\usepackage{amsmath}
\usepackage{amsfonts}
\usepackage{amssymb,dsfont}
\usepackage{graphicx}
\usepackage{amssymb}
\usepackage[vcentermath]{youngtab}
\usepackage[all]{xy}
\usepackage{pstricks}
\usepackage{dsfont}%
\usepackage{cases}
\usepackage{comment}

\usepackage{placeins}
\usepackage{xspace}
\usepackage{cancel} % to make the barred text notation

\usepackage{slashed}
\usepackage[caption=false, labelformat=simple, listofformat=subsimple, labelfont=default, margin=5pt,
    justification=raggedright]{subfig}

\makeatletter
	\renewcommand{\p@subfigure}{}
\makeatother

\usepackage{natbib}

\usepackage{feynmf}

\newcommand{\beq}{\begin{eqnarray}}
\newcommand{\eeq}{\end{eqnarray}}

\newcommand{\bmp}{\noindent\begin{minipage}{16cm}}
\newcommand{\emp}{\end{minipage}\vskip 7mm} % 7mm untightened

    \newcommand{\ii}{\mathrm{i}}
    \newcommand{\dd}{\mathrm{d}}
    
    \newcommand{\Pf}{\mathrm{Pf}}

    \newcommand{\SU}{\mathrm{SU}} 
    \newcommand{\Sp}{\mathrm{Sp}}

\def\lsim{\mathrel{\rlap{\lower4pt\hbox{\hskip1pt$\sim$}}
    \raise1pt\hbox{$<$}}}                % less than or approx. symbol
\def\gsim{\mathrel{\rlap{\lower4pt\hbox{\hskip1pt$\sim$}}
    \raise1pt\hbox{$>$}}}                % greater than or approx. symbol

\begin{document}
%%%%%%%%%%%%%%%%%%%%%%%%%%%%%%%%%%%%%%%%%%%%%%%%%%%%%%%%%%%%%%%%%%%%%%%%%%%

\title{\texorpdfstring{\Large\color{rossoCP3}  Elementary Goldstone Higgs boson and dark matter}{Elementary Goldstone Higgs boson and dark matter}}
\author{Tommi {\sc Alanne}}
\email{tommi.alanne@jyu.fi}
\affiliation{Department of Physics, University of Jyv\"askyl\"a, \\
                      P.O.Box 35 (YFL), FI-40014 University of Jyv\"askyl\"a, Finland}
%\affiliation{Helsinki Institute of Physics, \\
%                      P.O.Box 64, FI-00014 University of Helsinki, Finland}
\author{Helene {\sc Gertov}}
\email{gertov@cp3.dias.sdu.dk} 
\affiliation{{\color{rossoCP3} {CP}$^{ \bf 3}${-Origins}} \& the Danish Institute for Advanced Study {\color{rossoCP3}\rm{Danish IAS}},  University of Southern Denmark, Campusvej 55, DK-5230 Odense M, Denmark.}
\author{Francesco {\sc Sannino}}
\email{sannino@cp3.dias.sdu.dk}
\affiliation{{\color{rossoCP3} {CP}$^{ \bf 3}${-Origins}} \& the Danish Institute for Advanced Study {\color{rossoCP3}\rm{Danish IAS}},  University of Southern Denmark, Campusvej 55, DK-5230 Odense M, Denmark.}
\author{Kimmo {\sc Tuominen}}
\email{kimmo.i.tuominen@helsinki.fi}
\affiliation{Department of Physics, University of Helsinki, 
%                      P.O.Box 64, FI-00014 University of Helsinki, Finland}                 
\& Helsinki Institute of Physics, \\
                      P.O.Box 64, FI-00014 University of Helsinki, Finland}
\begin{abstract}
We investigate a perturbative extension of the Standard Model featuring elementary pseudo-Goldstone Higgs and dark matter particles. These are two of the five Goldstone bosons parametrising the $\SU(4)/\Sp(4)$ coset space.  They acquire masses, and therefore become pseudo-Goldstone bosons, due to the embedding of the Yukawa  and the electroweak gauge interactions that do not preserve the full $\SU(4)$ symmetry.  At the one-loop order the top corrections dominate and align the vacuum in the direction where the Higgs is mostly a pseudo-Goldstone boson. Because of the perturbative and elementary nature of the theory,  the quantum corrections are precisely calculable. The remaining pseudo-Goldstone boson is identified with the dark matter candidate because it is neutral with respect to the Standard Model and stable.   By a direct comparison with the Large Hadron Collider experiments, the model is found to be phenomenologically viable. Furthermore the  dark matter particle leads to the observed thermal relic density while respecting the most stringent current experimental constraints. 
  \\
[.1cm]
{\footnotesize  \it Preprint: CP$^3$-Origins-2014-040 DNRF90 \& DIAS-2014-40 \& HIP-2014-29/TH}
\end{abstract}
\maketitle
\newpage
 
\section{Introduction}

    The discovery of the Higgs boson constitutes a tremendous success for the Standard Model (SM) of particle interactions. However, the SM fails to explain, among other things, the missing mass in the universe. A solution to this problem requires extending the SM. Theoretically, it is appealing to analyse extensions featuring larger symmetries.  
    A time-honoured example is supersymmetry that, however, requires a very large number of, so far, unobserved particles. 
    
 Furthermore, it is a fact that the Higgs mass is an order of magnitude lighter than the natural scale for electroweak symmetry breaking which is around $4\pi v$ with $v\simeq 246$~GeV. This could be an accidental feature or more interestingly, it could emerge because of an underlying near-conformal dynamics of either composite  \cite{Sannino:1999qe,Sannino:2004qp,Dietrich:2006cm,Foadi:2012bb,Fodor:2014pqa}  
 or  perturbative 
 \cite{Grinstein:2011dq, Antipin:2011aa,Antipin:2012kc,Antipin:2012sm,Antipin:2013exa} nature. It has also been speculated that the Higgs could emerge as a composite pseudo-Goldstone boson (pGB) from extensions of the dynamical electroweak symmetry breaking paradigm \cite{Kaplan:1983fs,Kaplan:1983sm}. 

A unified description, both at the effective and fundamental Lagrangian level, of models of composite (Goldstone) Higgs dynamics was recently provided in \cite{Cacciapaglia:2014uja}. We refer to \cite{Cacciapaglia:2014uja} also for a critical discussion of the main differences, similarities, benefits and shortcomings of the different ways one can realize a composite nature of the electroweak sector of the SM.  There the minimal underlying realization in terms of fundamental strongly coupled gauge theories supporting the global symmetry breaking pattern SU(4)/Sp(4) $\sim$ SO(6)/SO(5) was also discussed. The realisation consists of an SU(2) gauge theory with two Dirac fermions transforming according to the fundamental representation of the gauge group. Thanks to this identification  it has been possible  \cite{Cacciapaglia:2014uja} to adapt first principle lattice results \cite{Lewis:2011zb,Hietanen:2013fya,Hietanen:2014xca} to make important predictions for the massive spectrum of models of composite (Goldstone) Higgs. The results have an immediate impact on guiding experimental searches of new physics at the LHC \footnote{Historically the symmetry breaking scenario SO(6)$\sim$SU(4)$\to$Sp(4)$\sim$SO(5)  were introduced for (ultra) minimal Technicolor models \cite{Appelquist:1999dq,Duan:2000dy,Ryttov:2008xe} while the composite GB Higgs example was discussed in \cite{Katz:2005au,Gripaios:2009pe,Galloway:2010bp,Barnard:2013zea,Ferretti:2013kya}. The discussion of possible ultraviolet completions for the Little Higgs models appeared in \cite{Batra:2007iz}.}. 
 
 In this paper we take a complementary route by considering a fully perturbative SM extension in which an {\it elementary pGB} rather than a composite Goldstone Higgs naturally emerges from the precisely calculable dynamics. Differently from the composite case we also have an elementary pGB dark matter (DM) candidate. The reason why such a state cannot be stable in the composite model is that the underlying fermionic dynamics induces a Wess-Zumino-Witten term~\cite{Wess:1971yu,Witten:1983ar,Witten:1983tw} that efficiently mediates its decay into SM particles. This operator is absent in a renormalizable  Lagrangian of elementary scalars.   
  
Here we therefore extend  the work in \cite{Cacciapaglia:2014uja} to a perturbative renormalizable description. The elementary template uses the $\SU(4)$ global symmetry for the Higgs sector   that breaks spontaneously to $\Sp(4)$.   Among the salient features of this extension we mention: 
\begin{itemize}
\item The elementary Higgs sector is perturbative and the quantum corrections can be computed in controllable manner. This allows a reliable determination of the ground state, quantum corrections and spectrum of the theory.

\item The SM fermions acquire mass without the need of invoking  unknown dynamics \cite{Ferretti:2013kya}. 

\item The top corrections naturally lead to a pGB nature of the elementary Higgs and DM particles\footnote{This is very different from the composite case. The reason is that the theory here is renormalizable  and therefore there is no cutoff-induced dependence for the quantum corrected potential of the theory. In turn this also means that no new sources, from yet unknown dynamics, are needed to realise a pGB Higgs theory. 
 }. 

\item{The new heavy scalar spectrum appears in the (multi) TeV range.}

\item{The theory predicts very small deviations with respect to the SM Higgs couplings with the exception of a highly suppressed trilinear Higgs coupling.}

\item{It is possible to obtain the observed DM thermal relic density while satisfying the most stringent current experimental constraints. }
\end{itemize}

The paper is organised as follows. In Section~\ref{sec:scalar} we introduce the renormalizable Lagrangian, determine its ground state and the associated tree-level  scalar spectrum. In Section~\ref{sec:SU4br} we embed the electroweak interactions, give masses to the SM fermions, most notably the top quark, and add sources of explicit $\SU(4)$ breaking. The construction and general features of the one-loop quantum corrected potential are presented in \ref{sec:1loop}. The numerical study of the vacuum alignment, the spectrum and the phenomenological viability of the theory is considered in Section \ref{sec:LHC} where we also compare with collider constraints. Finally, we discuss the DM phenomenology in \ref{sec:DM}. We conclude in section~\ref{sec:concl}.

The appendices provide further details on the analysis: In Appendix~\ref{LFstability} we provide details about the determination of the ground state and tree-level stability of the Higgs potential. A realisation of the generators is given in Appendix~\ref{appgenerators} while the explicit definition of the Pfaffian is in Appendix~\ref{pfaffians}.

\section{Renormalizable potential with $\SU(4)$ to $\Sp(4)$ breaking }
    \label{sec:scalar}

The Higgs sector of the SM, before gauging the electroweak symmetry, has an $\SU_{\mathrm{L}}(2) \times \SU_{\mathrm{R}}(2)$ global symmetry.  These SM Higgs fields can be assembled in a matrix transforming as a bifundamental of this symmetry. This minimal realisation counts one scalar state, the Higgs, and three pions which upon spontaneous symmetry breaking to $\SU_{\mathrm{V}}(2)$ and electroweak gauging become the longitudinal components of the heavy gauge bosons in the unitary gauge.

However, in this elegant and minimal realization the Higgs mass operator is not protected by any symmetry and there is no room for any DM candidate. It is therefore interesting, as explained in the introduction, to explore the possibility to extend the symmetries of the Higgs sector to accommodate simultaneously the protection of the Higgs mass and the natural emergence of a DM candidate. 

In other words, we would like to generate a pGB Higgs and a pGB DM candidate in a fairly minimal setup. 
This motivates us to use  $\SU(4)$  as a new enlarged symmetry with the Higgs matrix transforming according to the two-index antisymmetric representation.
In fact, when $\SU(4)$ breaks spontaneously to $\Sp(4)$, we have %only 
five Goldstones, three of which to become the longitudinal components of the weak gauge bosons. This scenario is realised by introducing the matrix of fields
    \begin{equation}
	\label{eq:Mdef}
	M=\left[\frac{\sigma+\ii\Theta}{2}+\sqrt{2}(\ii\Pi_i+\tilde{\Pi}_i)X^i\right]E,
    \end{equation}
    where $E$ is an antisymmetric matrix and $X^i$ are the hermitian matrices corresponding to the broken generators of $\SU(4)$ for the vacuum along $E$. The $\Pi_i$ fields thus correspond to the Goldstone bosons associated with the spontaneous breaking of $\SU(4)$ to $\Sp(4)$. Summation over repeated indices is always implied unless otherwise stated. The most general $\SU(4)$-symmetric renormalizable potential for $M$ is
    \begin{equation}
	\label{eq:pot}
	\begin{split}
	    V_M=&\frac{1}{2}m_M^2\mathrm{Tr}[M^{\dagger} M]+\left( c_M\Pf(M)+\mathrm{h.c.}\right)\\
		&+\frac{\lambda}{4}\mathrm{Tr}[M^{\dagger}M]^2+\lambda_1\mathrm{Tr}[M^{\dagger}MM^{\dagger}M]
		    -2\left(\lambda_2\Pf(M)^2+\mathrm{h.c.}\right)\\
		&+\left(\frac{\lambda_3}{2}\mathrm{Tr}[M^{\dagger}M]\Pf(M)+\mathrm{h.c.}\right),
	\end{split}
    \end{equation}
    where the coefficients $c_M, \lambda_2$, and $\lambda_3$ can, in principle, be complex, whereas 
    $m_M^2, \lambda$, and $\lambda_1$ are real. 
    Note that without the Pfaffian terms, the potential is actually symmetric under the full $\mathrm{U}(4)$ group instead of just $\SU(4)$. 
	
    The tree-level stability of this potential at large values of the fields has been studied in detail in Appendix \ref{LFstability}  resulting in the following sufficient constraint on the available parameter space of the scalar couplings: 
    \begin{equation}
	\label{eq:VacStab}
	\lambda+\Delta_{1}\lambda_1-|\lambda_{2R}|-|\lambda_{2I}|-|\lambda_{3R}|-|\lambda_{3I}|\geq 0,
    \end{equation}
    with  
    \begin{equation}
	\label{eq:VacStabDeltas}
	\Delta_1=\left\{\begin{array}{l}1,\quad\text{if }\lambda_1\geq0\\2,\quad\text{if }\lambda_1<0 \end{array}\right. ,
    \end{equation}
    and the subscripts $R$ and $I$ referring to the real and imaginary parts of the couplings, respectively.
 
    \subsection{Ground state}
	The vacuum structure of the potential has been investigated  in Appendix~\ref{LFstability}. For simplicity, we consider real couplings, i.e. we set $c_{MI}=\lambda_{2I}=\lambda_{3I}=0$ and consequently drop $R$ from the subscripts. 
We discover the following two minima of the potential:
	\begin{equation}
	    \label{eq:critPRe1}
	    \langle \sigma^2+\mathbf{\Pi}^2 \rangle=\frac{c_M-m_M^2}{\lambda+\lambda_1-\lambda_2 -\lambda_3},\qquad 
		\langle\Theta^2\rangle=\langle\mathbf{\tilde{\Pi}}^2\rangle=0,
	\end{equation}
	and 
	\begin{equation}
	    \label{eq:critPRe2}
	    \langle \Theta^2+\mathbf{\tilde{\Pi}}^2\rangle=\frac{-c_M-m_M^2}{\lambda+\lambda_1-\lambda_2 +\lambda_3},\qquad
		\langle\sigma^2\rangle=\langle\mathbf{\Pi}^2\rangle=0,
	\end{equation}
	where $\mathbf{\Pi}^2=\Pi_i\Pi_i$ and $\mathbf{\tilde{\Pi}}^2=\tilde{\Pi}_i\tilde{\Pi}_i$. 
	We choose the parameters of the potential such that the first minimum, Eq.~\eqref{eq:critPRe1}, is the global one. 

	The value of the potential at the global minimum then reads
	\begin{equation}
	    \label{eq:Vmin}
	      \langle V_M \rangle=-\frac{1}{4}(\lambda+\lambda_1-\lambda_2
		 -\lambda_3)  \langle \sigma^2+\mathbf{\Pi}^2 \rangle^2.
	\end{equation}

	  In both cases the pattern of spontaneous symmetry breaking  is  
	$\SU(4)$ to $\Sp(4)$\footnote{The reason is that the five broken generators, 
	collectively denoted with $X$,  have the property 
	$X^T = EXE^T$, and therefore $XE$ is also antisymmetric \cite{Duan:2000dy}. The ten unbroken generators, labelled 
	collectively by $S$, that leave the vacuum invariant satisfy the condition $S E + ES^T = 0$.}.
	Further details can be found in Appendix \ref{LFstability}.

    \subsection{Spectrum} 
Thanks to the $\SU(4)$ invariance we can choose the ground state along $E$, meaning that we set to zero the expectation values of all the fields except $\sigma$ in \eqref{eq:critPRe1}.  	
We can now determine	the masses of the $\sigma$ and $\Theta$ particles that read
	\begin{equation}
	    \label{eq:masssigmatheta}
	    \begin{split}
		m^2_{\sigma}&=2(c_M-m_M^2),\\
		m_{\Theta}^2&
		    =\frac{c_M(2\lambda+2\lambda_1+2\lambda_2-\lambda_3)
		    -m_M^2(4\lambda_2+\lambda_3)}{\lambda+\lambda_1-\lambda_{2} -\lambda_3}.
	    \end{split}
	\end{equation}
The scalar partners of the Goldstone bosons $\Pi_i$, i.e. the $\tilde\Pi_i$ fields, form a quintuplet of 
	$\Sp(4)$ with mass: 
	\begin{equation}
	    \label{eq:mPiT}
	    m_{\tilde\Pi}^2=\frac{c_{M}(2\lambda+4\lambda_1-\lambda_3)
		-m_M^2(2\lambda_1+2\lambda_{2}+\lambda_3)}{\lambda+\lambda_1-\lambda_{2}
		 -\lambda_3}.
	\end{equation}

\section{Weak interactions, fermion masses and explicit breaking of SU(4)}
    \label{sec:SU4br}

    \subsection{The electroweak embedding}
There can be several ways to embed the electroweak gauge group   
$\SU(2)_{\mathrm{L}}\times \mathrm{U}(1)_Y$
into the larger group $\SU(4)$.
In particular, we
	are interested in having the possibility that the entire Higgs-doublet could be identified with four of the five 
	Goldstones of the theory. We know that this is possible as shown \cite{Katz:2005au,Gripaios:2009pe,Galloway:2010bp,Barnard:2013zea,Ferretti:2013kya} and recently investigated in a scenario of
	unified composite Higgs dynamics of Goldstone and traditional Technicolor kind in \cite{Cacciapaglia:2014uja}. The choice of 
	the embedding is not entirely arbitrary because the dynamics itself coming from the electroweak as well as other 
	sectors determines coherently the embedding scenario. Following~\cite{Galloway:2010bp}, we embed the full 
	custodial symmetry group of the SM, $\SU(2)_{\mathrm{L}}\times \SU(2)_{\mathrm{R}}$, in $\SU(4)$
	by identifying the left and right generators, respectively, with
	\begin{equation}
	    \label{eq:gensCust}
	    T^i_{\mathrm{L}}=\frac{1}{2}\left(\begin{array}{cc}\sigma_i & 0 \\ 0 & 0\end{array}\right),\quad\text{and}\quad
	    T^i_{\mathrm{R}}=\frac{1}{2}\left(\begin{array}{cc} 0 & 0 \\ 0 & -\sigma_i^{T}\end{array}\right),
	\end{equation}
	where $\sigma_i$ are the Pauli matrices. The generator of the hypercharge is then identified with the third generator 
	of the $\SU(2)_{\mathrm{R}}$ group, $T_Y=T^3_{\mathrm{R}}$.

	Considering the above embedding of the electroweak sector, we want to first identify the vacua that leave the 
	electroweak symmetry intact. 
	As discussed in~\cite{Galloway:2010bp}, there are two inequivalent vacua of this kind,
	\beq
	E_A = \left( \begin{array}{cc}
	\ii \sigma_2 & 0 \\
	0 & \ii \sigma_2
	\end{array} \right)\,, \qquad E_B = \left( \begin{array}{cc}
	\ii \sigma_2 & 0 \\
	0 & - \ii \sigma_2
	\end{array} \right)\,,
	\eeq
	and by inequivalent one means that 
	they cannot be related to one  another by an $\SU(2)_{\mathrm{L}}$ transformation.
	
	We chose the normalization to be real.
	With either choice, the physical properties of the pGBs are the same:
	in~\cite{Gripaios:2009pe}, the authors consider $E_A$ to build their model, while in~\cite{Galloway:2010bp} the model 
	is constructed around $E_B$. In this paper, we will use $E_B$.

	There is another alignment of the condensate which is of physical interest, given by the matrix
	\beq
	E_H  =\left( \begin{array}{cc}
	0 & 1 \\
	-1 & 0
	\end{array} \right) \ .
	\eeq
	This vacuum completely breaks the electroweak symmetry, and can therefore be used to construct a Technicolor
	model~\cite{Appelquist:1999dq,Duan:2000dy,Ryttov:2008xe}.

	We start by defining the vacuum of the theory as a superposition of the two vacua defined above,
	\beq
	E_{\theta} =\cos \theta\; E_B + \sin \theta\; E_H\, ,
	\eeq
	such that $E_{\theta}^{\dagger} E_{\theta}^{\vphantom{\dagger}} = 1$.
	The angle $\theta$ is, at this stage, a free parameter, which interpolates between a model with completely broken
	EW symmetry when $\theta=\pi/2$ and an unbroken phase for $\theta=0$.
	
	To implement the above vacuum structure, we reparameterize the scalar matrix as
	\begin{equation}
	    \label{eq:M_EW}
	    M=\left[\frac{\sigma+\ii\Theta}{2}+\sqrt{2}(\ii\Pi_i+\tilde{\Pi}_i)X_{\theta}^i\right]E_{\theta},
	\end{equation}
	where the generators $X_{\theta}$ are associated with the vacuum $E_{\theta}$ and are given in the 
	Appendix~\ref{appgenerators}. 
	The spontaneous breaking of $\SU(4)$ to $\Sp(4)$ occurs when $M$ acquires the 
	 vacuum expectation value
	\begin{equation}
	    \label{eq:Mvev}
	    \langle M\rangle=\frac{\langle\sigma\rangle}{2}E_{\theta}=\frac{v}{2}E_{\theta}\, , 
	\end{equation}
	with  $v$ given by Eq.~\eqref{eq:critPRe1},
	\begin{equation}
	    \label{eq:vev}
	    v^2=\frac{c_M-m_M^2}{\lambda+\lambda_1-\lambda_2 -\lambda_3}.
	\end{equation}

	We gauge the electroweak group by defining the covariant derivative of $M$,
	\begin{equation}
	    \label{eq:covM}
	    D_{\mu}M=\partial_{\mu}M-\ii\left(G_{\mu}M+MG_{\mu}^{\mathrm{T}}\right).
	\end{equation}
	The gauge field, $G_{\mu}$, can be written as
	\begin{equation}
	    \label{eq:gaugeG}
	    G_{\mu}=gW^i_{\mu}T_{\mathrm{L}}^i+g'B_{\mu}T^3_{\mathrm{R}}\, ,
	\end{equation}
	where the generators $T^i_{\mathrm{L}}$ and $T^3_{\mathrm{R}}$ are given by Eq.~\eqref{eq:gensCust}. 
	We, thus, upgrade the kinetic term for $M$ to be electroweak gauge invariant,
	\begin{equation}
	    \label{eq:kinM}
	    \frac{1}{2}\mathrm{Tr}\left[D_{\mu}M^{\dagger}D^{\mu}M\right].
	\end{equation}
When $M$ acquires a vacuum expectation value, the weak gauge bosons acquire masses depending on the value of $\theta$, i.e. 
	\begin{equation}
	    \label{eq:WBosMasses}
	    m_W^2=\frac{1}{4}g^2v^2\sin^2\theta, \quad\text{and}\quad m_Z^2=\frac{1}{4}(g^2+g'^2)v^2\sin^2\theta.
	\end{equation}  

    \subsection{Top quark mass and interactions }
    Due to the fact that we are assuming $M$ to be constituted by elementary (pseudo) scalars we are entitled to introduce the operators responsible to give mass to the SM fermions. Here we  concentrate on the top quark term.  
	 To this end, we define projectors $P_1$ and $P_2$ ~\cite{Galloway:2010bp}
	that pick the components of the weak doublet contained in $M$. 
	\begin{equation}
	    \label{eq:SU2projectors}
	    P_1=\frac{1}{\sqrt{2}}\left(
	    \begin{array}{cccc}
		0	&   0	&   1	&   0\\
		0	&   0	&   0	&   0\\
		-1	&   0	&   0	&   0\\
		0	&   0	&   0	&   0
	    \end{array}
	    \right),\qquad
	    P_2=\frac{1}{\sqrt{2}}\left(
	    \begin{array}{cccc}
		0	&   0	&   0	&   0\\
		0	&   0	&   1	&   0\\
		0	&   -1	&   0	&   0\\
		0	&   0	&   0	&   0
	\end{array}
	\right).
	\end{equation}
	In terms of these projectors, the Yukawa term for the top reads
	\begin{equation}
	    \label{eq:topYuk}
	    \mathcal{L}_{\mathrm{Yuk}}=y_t(Qt^{c})^{\dagger}_{\alpha}\mathrm{Tr}[P_{\alpha}M]+\mathrm{h.c.},
	\end{equation}
	where $\alpha$ is the $\SU(2)_{\mathrm{L}}$ index. The top quark then acquires the following mass:
	\begin{equation}
	    \label{eq:mtop}
	    m_t=\frac{y_t}{\sqrt{2}}v\sin\theta.
	\end{equation}

    \subsection{An explicit breaking of $\SU(4)$}
    
    There can be several renormalizable terms able to break the $\SU(4)$ symmetry explicitly. However, since we have in mind $\Pi_5$ to be the potential DM candidate we add to the model potential a term\footnote{Interestingly, after radiative corrections to the mass of $\Pi_5$ this term is also required to ensure the overall quantum stability of the theory.} 
    \begin{equation}
	    \label{eq:muMterm}
	    V_{\mathrm{br}}=\frac{1}{8}\mu_M^2\mathrm{Tr}\left[E_A M\right]\mathrm{Tr}\left[E_A M\right]^*,
	\end{equation}
	where
	\begin{equation}
	    \label{eq:EA}
	    E_A = \left( 
	    \begin{array}{cc}
		\ii\sigma_2 & 0 \\
		0 & \ii\sigma_2
	    \end{array} \right)\, .   
	\end{equation}
This term preserves a $Z_2$ symmetry 
	under which $\Pi_5\rightarrow -\Pi_5,\ \tilde{\Pi}_5\rightarrow -\tilde{\Pi}_5$. This ensures that $\Pi_5$ is an absolutely stable electroweak neutral particle. It is straightforward to see that in the field components, the breaking term becomes 
	\begin{equation}
	    \label{eq:muMtermComp}
	    V_{\mathrm{br}}=\frac{1}{2}\mu_M^2\left[(\Pi_5)^2+(\tilde{\Pi}_5)^2\right].
	\end{equation}

\section{Quantum potential}
    \label{sec:1loop}
  Having set up the model, we are ready to determine its quantum effective potential. Since the electroweak and top sectors do not preserve the full symmetries of the classical vacuum, they will affect the vacuum structure.

At the one-loop level the effective potential reads
	    \begin{equation}
		\label{eq:deltaV}
		\delta V(\Phi)=\frac{1}{64\pi^2}\mathrm{Str}\left[{\cal M}^4(\Phi)\left(\log\frac{{\cal M}^2(\Phi)}
		    {\mu_0^2}-C\right)\right]+V_{\mathrm{GB}},
	    \end{equation}
	    where ${\cal M}(\Phi)$ is the tree-level mass matrix for the background value of the matrix of fields, $M$, that we denote by $\Phi$, and the supertrace, $\mathrm{Str}$, is defined by
\begin{equation}
\mathrm{Str} = \sum_{\text{scalars}}-2\sum_{\text{fermions}}+3\sum_{\text{vectors}}.
\end{equation}
We have $\displaystyle{C=\frac{3}{2}}$ for scalars and fermions, while $C=\displaystyle{\frac{5}{6}}$ for the gauge bosons. Here $V_{\mathrm{GB}}$ contains the GB contributions and $\mu_0$ is a reference renormalization scale.  We have already added the appropriate counter terms to cancel the ultraviolet divergences using dimensional regularization in the $\overline{\mathrm{MS}}$ scheme\footnote{Treating the Goldstone boson corrections to the potential as done for the massive scalars would lead to infrared divergences due to their vanishing masses. There are several ways of dealing with this issues, for example adding some characteristic mass scale as an infrared regulator. However, since the massive scalars give the dominant contribution to the vacuum structure of the theory, we simply neglect the GB contributions.}. 

In the numerical calculation we consider the full effective potential (\ref{eq:deltaV}) as detailed above. For concreteness and illustration, let us here write down the one-loop correction to the potential  along the $\sigma$ direction. To reduce the number of unknowns, we 
consider the limit when all the non-Goldstone scalars have equal tree-level masses before adding explicit $\SU(4)$ breaking terms: 
\begin{equation}
    \label{eq:equalMass}
    {\cal M}_{\sigma}^2(v)={\cal M}_{\Theta}^2(v)={\cal M}_{\tilde{\Pi}}^2(v)=M_S^2. 
\end{equation}
After adding the explicit $\SU(4)$ breaking term of Eq.~\eqref{eq:muMtermComp}, the mass terms for $\Pi_5$ and $\tilde{\Pi}_5$ fields read
\begin{equation}
    \label{eq:Pi5masses}
    {\cal M}_{\tilde{\Pi}_5}^2(v)=M_S^2+\mu_M^2,\quad {\cal M}_{\Pi_5}^2(v)=\mu_M^2.
\end{equation}
Then the background dependent masses are
\begin{equation}
    \label{eq:BGmasses}
    \begin{split}
	&{\cal M}_{\sigma}^2(\sigma)=\frac{1}{2}M_S^2\left(\frac{3\sigma^2}{v^2}-1\right),
	    \quad {\cal M}_{\Theta}^2(\sigma)={\cal M}_{\tilde{\Pi}_{1,2,3,4}}^2(\sigma)=
	      M_S^2+\tilde{\lambda}(\sigma^2-v^2),\\
	&{\cal M}_{\tilde{\Pi}_{5}}^2(\sigma)=M_S^2+\mu_M^2+\tilde{\lambda}(\sigma^2-v^2),\quad
	    {\cal M}_{\Pi_{5}}^2(\sigma)=\mu_M^2+\frac{1}{2}M_S^2\left(\frac{\sigma^2}{v^2}-1\right),
    \end{split}
\end{equation}
where $\tilde{\lambda}=\lambda+4\lambda_1$.

The one loop potential in the scalar sector reads
\begin{equation}
    \label{eq:corrSc}
    \begin{split}
    \delta V_{\mathrm{sc}}(\sigma)=\frac{1}{64\pi^2}&\left[\frac{M_S^4(3\sigma^2-v^2)^2}{4v^4}\left(
	\log\frac{M_S^2(3\sigma^2-v^2)}{2v^2\mu_0^2}-\frac{3}{2}\right)\right.\\
	&+5\left(M_S^2+\tilde{\lambda}(\sigma^2-v^2)\right)^2
	\left(\log\frac{M_S^2+\tilde{\lambda}(\sigma^2-v^2)}{\mu_0^2}-\frac{3}{2}\right)\\
	&+\left(M_S^2+\mu_M^2+\tilde{\lambda}(\sigma^2-v^2)\right)^2
	\left(\log\frac{M_S^2+\mu_M^2+\tilde{\lambda}(\sigma^2-v^2)}{\mu_0^2}-\frac{3}{2}\right)\\
	&+\left.\frac{\left(2v^2\mu_M^2+M_S^2(\sigma^2-v^2)\right)^2}{4v^4}\left(
	\log\frac{2v^2\mu_M^2+M_S^2(\sigma^2-v^2)}{2v^2\mu_0^2}-\frac{3}{2}\right)\right].
    \end{split}
\end{equation}

Similarly, the corrections from the electroweak and top sectors are
\begin{align}
    \label{eq:corrEWtop}
    \begin{split}
        \delta V_{\mathrm{EW}}(\sigma)=&\frac{3}{1024\pi^2}\sigma^4\sin^4\theta\left[2g^4\left(
	    \log\frac{g^2\sigma^2\sin^2\theta}{4\mu_0^2}-\frac{5}{6}\right)\right.\\
	&\left.\qquad\qquad\qquad\quad+(g^2+g^{\prime\, 2})^2\left(\log\frac{(g^2+g^{\prime\, 2})\sigma^2\sin^2\theta}{4\mu_0^2}
	    -\frac{5}{6}\right)\right]
    \end{split}\\
        \delta V_{\mathrm{top}}(\sigma)=&-\frac{3}{64\pi^2}\sigma^4\sin^4\theta y_t^4\left(
	    \log\frac{y_t^2\sigma^2\sin^2\theta}{2\mu_0^2}-\frac{3}{2}\right).
\end{align}

	    Moreover, we trade the renormalization scale, $\mu_0$, with the vacuum expectation value in the 
	    $\sigma$ direction by requiring the cancellation of the tadpoles via: 
	    \begin{equation}
		\label{eq:notads}
		\left.\frac{\partial\delta V(\sigma)}{\partial\sigma}\right|_{v}=0.
	    \end{equation}
	    This yields
	    \begin{equation}
		\label{eq:logmu0}
		\log\mu_0^2=\left.\frac{\frac{\partial}{\partial\sigma}\mathrm{Str}\left[M^4(\sigma)\left(\log M^2(\sigma)
		    -C\right)\right]}{\frac{\partial}{\partial\sigma}\mathrm{Str}\left[M^4(\sigma)\right]}\right|_v \ .
	    \end{equation}
	    Substituting this expression into the one-loop potential we replace $\mu_0$ with an expression as function of $v$. 
	    
	     We still need to minimize the potential with respect to the embedding angle $\theta$
	    \begin{equation}
		\label{eq:SPtheta}
		\left.\frac{\partial\delta V(\sigma)}{\partial\theta}\right|_v=0.
	    \end{equation}
	
Before discussing the actual ground state of the theory,
we note that because the electroweak and top interactions do not preserve the original $\SU(4)$, the $\sigma$ and $\Pi_4$ states mix.   
	    Denoting the mass eigenstates with $h^0$ and $H^0$, we express $\sigma$ and $\Pi_4$ via the mixing angle $\alpha$ as follows
		\begin{equation}
		    \label{eq:sigmaPi4}
			    \left(\begin{array}{c}
			\sigma\\
			\Pi_4
		    \end{array}\right)
		    =
		    \left(\begin{array}{cc}
			\cos\alpha & -\sin\alpha\\
			\sin\alpha & \cos\alpha
		    \end{array}\right)
		    \left(\begin{array}{c}
			h^0\\
			H^0
		    \end{array}\right) \ .
		\end{equation}
	    Here, $h^0$ is the lightest eigenstate identified with the 125~GeV Higgs observed at the LHC. This situation mirrors what happens in the composite case \cite{Cacciapaglia:2014uja}.  
	    
	    It is instructive to consider the analytic dependence of $\alpha$ on $\theta$ and the top-Yukawa coupling. Although in the final numerical analysis, we do not drop either the dependence on the electroweak gauge couplings or the explicit $\SU(4)$ breaking terms, we do so here to keep the expression simple. In this limit, we obtain for the mixing angle:
\begin{equation}
    \label{eq:alpha}
    \alpha=\frac{\pi}{2}-\frac{6\theta^3v^4y_t^4}{3M_S^4+32\pi^2M_S^2v^2+24{\tilde{\lambda}}^2v^4}
	\left(3\log M_S^2-3\log\frac{y_t^2v^2\theta^2}{2}-2\right)+{\cal O}(\theta^5)\ .
\end{equation}
In the expression above we have also Taylor expanded in $\theta$. As expected, in the absence of the top Yukawa coupling, as well as the gauge and explicit breaking, the two states do not mix, and the lightest state is $h^0=\Pi_4$. 

We are ready to discuss the properties of the ground state of the theory. We start with the observation that only the electroweak and top corrections depend explicitly on the embedding angle $\theta$. Furthermore because the top-Yukawa interaction strength dominates over the electroweak ones we can concentrate, for a qualitative understanding, on the top contribution. For small values of the field $\sigma$ the log in Eq.~\eqref{eq:corrEWtop} is negative and therefore the top corrections privilege small values of $\theta$. This expectation is, indeed, confirmed by the detailed numerical analysis below.

\section{Numerical analysis and collider limits}
    \label{sec:LHC}
    	    In the numerical analysis we evaluate the full effective potential (\ref{eq:deltaV}), and work under the assumptions detailed in the previous section. The free parameters of the model are $M_S$, $\tilde{\lambda}$, $v$, $\theta$ and $\mu_M$. The value of $\mu_M$ has little effect on the vacuum structure and the Higgs mass when the explicit breaking is small, $\mu_M\ll v$, and we choose as a benchmark value $\mu_M=100$ GeV. First, to be in accordance with the experiments, we are searching for a solution that yields the correct $W,Z$ and top masses, i.e. for which $v\sin\theta=v_w=246$~GeV. This condition is satisfied on the red contour plotted in the panels of Fig.~\ref{fig:SP100}. Second, the solution has to correspond to the minimum of the potential. For fixed $M_S$ and $\tilde{\lambda}$ the stationarity condition with respect to the vacuum angle, 	    
	    Eq.~\eqref{eq:SPtheta}, provides a curve $v(\theta)$; this is the blue contour in Fig.~\ref{fig:SP100}.
	    Moreover, with corresponding parameter values, we plot in Fig.~\ref{fig:SP100} the 
	    regions for which the second derivative of the potential with respect to $\theta$ is positive. Thus, on these 
	    shaded regions the stationary point is a minimum.
	    In practice, we are interested  in the cases where the red and blue curves intersect 
	    on the shaded region, and at the intersection the value of the correct Higgs mass is reproduced.
	    This condition fixes the value of the tree-level scalar mass, $M_S$, and $\tilde{\lambda}$ remains as a free parameter.\footnote{The mass scale $M_S$ obtained in this way is large, of the order of few TeV. This can be understood as a consequence of the fact that essentially all of the Higgs mass is generated radiatively in this model.} In Fig.~\ref{fig:SP100} we have considered the values $\tilde{\lambda}=0.1$, $0.5$ and $1.5$. For larger values of the coupling $\tilde{\lambda}$ the situation is essentially similar to the one in the rightmost panel of Fig. \ref{fig:SP100}. The quantitative difference is that the value of $\theta$ determined by the intersection of the red and blue curves moves towards large values and the scale $v$ decreases compatibly with the fact that the Higgs is becoming less pGB-like scalar state.

	    \begin{figure}
		\begin{center}
		    \label{fig:SP100}
		    \includegraphics[width=0.3\textwidth]{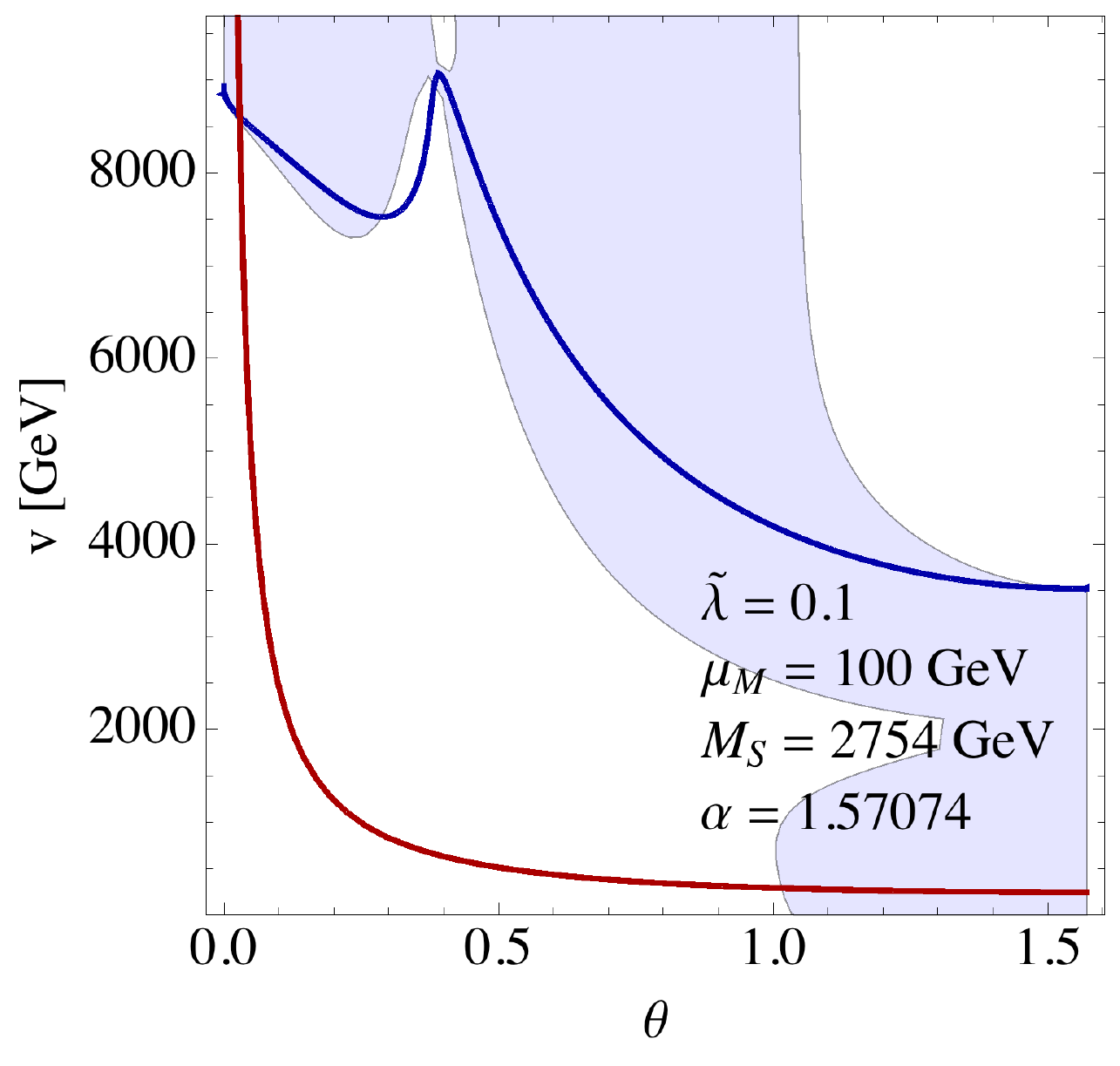}
		    \includegraphics[width=0.3\textwidth]{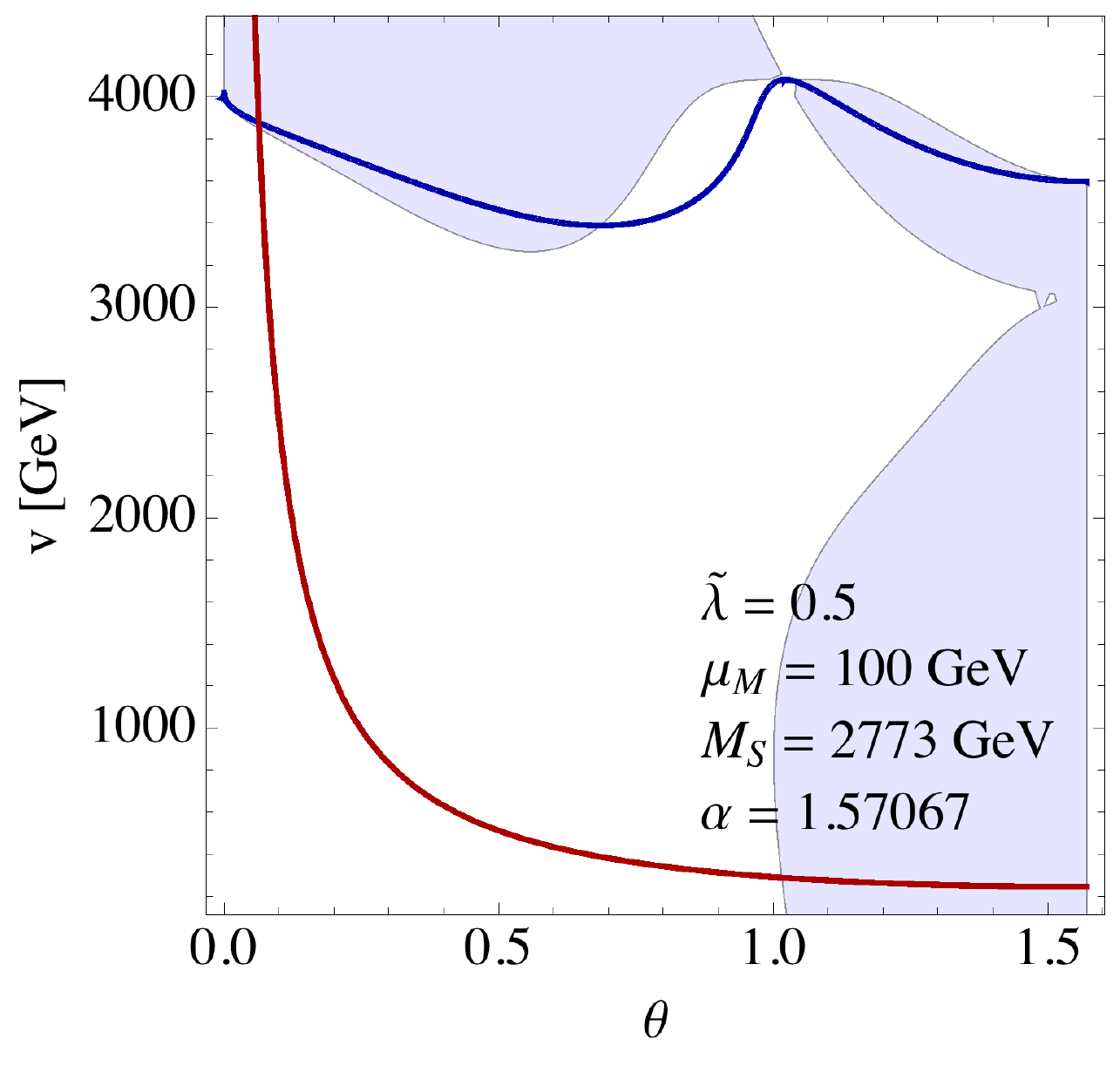}
		    \includegraphics[width=0.3\textwidth]{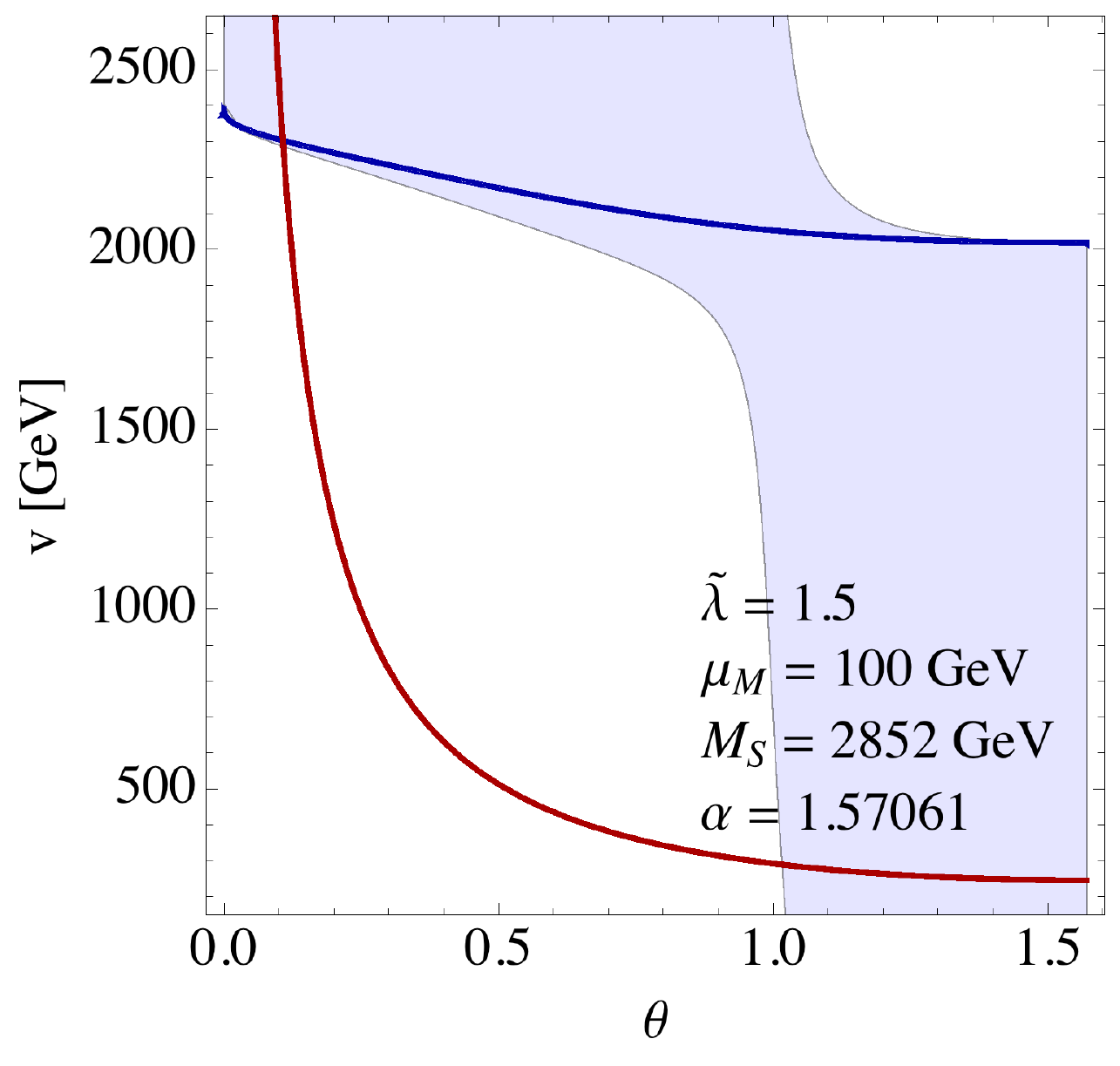}
	    \end{center}
	    \caption{The blue contour represents the stationary points with respect to $\theta$,
		the blue regions show where the second derivative with respect to $\theta$ is positive (stationary point
		contour on a blue region is thus a minimum) and the red contour shows the points that give the correct EW gauge boson 
		and top quark masses. The tree-level masses of all the heavy scalars are assumed to be of the same order, $M_S$, and its
		value is fixed by identifying lightest eigenstate of the $\sigma-\Pi_4$ mixing with the observed 125 GeV scalar.}
		\label{fig:SP100}
	    \end{figure}

     	    Although in the analysis of the one-loop effects, the light SM fermions play a negligible role, they are relevant when considering the collider or DM phenomenology. Thus, we couple them to $M$ following the way we added the top interactions which means:
	    \begin{equation}
		\label{eq:upYuk}
		\mathcal{L}_{\mathrm{Yuk}}^{\mathrm{up}}=y_u(Qu^{c})^{\dagger}_{\alpha}\mathrm{Tr}[P_{\alpha}M]+\mathrm{h.c.},
	    \end{equation}
	    for up-type fermions, and 
	    \begin{equation}
		\label{eq:downYuk}
		\mathcal{L}_{\mathrm{Yuk}}^{\mathrm{down}}=y_d(Qd^{c})^{\dagger}_{\alpha}\mathrm{Tr}[-\ii 
		    (\sigma_2)_{\alpha\beta} P_{\beta}M^*]+\mathrm{h.c.},
	    \end{equation}
	    for down-type fermions. 
	    
	    Then the Higgs self-couplings and the couplings of the Higgs with the gauge and fermion sectors read
	    \begin{equation}
		\label{eq:couplings}
		\begin{split}
		    &\lambda_{hhhh}=6\lambda_{\mathrm{eff}},\quad
			\lambda_{hhh}=6\lambda_{\mathrm{eff}}\,v\cos\alpha\\
		    &g_{hWW}=\frac{1}{2}g^2v\sin\theta\sin(\theta+\alpha),\quad 
			g_{hZZ}=\frac{1}{2}(g^2+g^{\prime\, 2})v\sin\theta\sin(\theta+\alpha)\\
		    &y_{hff}=\frac{y_f}{\sqrt{2}}\sin(\theta+\alpha),
		\end{split}
	    \end{equation}
	    where $\lambda_{\mathrm{eff}}=\lambda+\lambda_1-\lambda_{2} -\lambda_3$.

	    The difference compared with the SM couplings for the gauge and fermion sectors can be parameterized by 
	    two coefficients, 
	    \begin{equation}
		\label{eq:cVcf}
		c_V=\frac{g_{hVV}}{g_{hVV}^{\mathrm{SM}}}=\frac{v\sin\theta\sin(\theta+\alpha)}{v_w}\quad\text{and}\quad
	    	c_f=\frac{y_{hff}}{y_{hff}^{\mathrm{SM}}}=\sin(\theta+\alpha).
	    \end{equation}
	    We have checked numerically that even for very large values $\tilde{\lambda}\leq10$,
	    the vector and fermion couplings differ from the SM values by less than 3\%. 
	    This is in good agreement with the current experimental
	    bounds from the CMS experiment~\cite{CMS:2014ega}
	    \begin{equation}
		\label{eq:CMSbounds}
		c_V=1.01^{+0.07}_{-0.07},\quad\text{and}\quad c_f=0.89^{+0.14}_{-0.13},
	    \end{equation}
	    and especially for small $\tilde{\lambda}$ these couplings are almost indistinguishable from the SM ones.

	    However, the Higgs self couplings deviate more from their SM counterparts. Especially, since the mixing $\alpha$ is nearly $\pi/2$, the Higgs trilinear coupling is highly suppressed compared 
	    to the SM value as it is clear by its expression	    
	    \begin{equation}
		\label{eq:Lhhh}
		\lambda_{hhh}=\frac{3M_S^2\cos\alpha}{v}
	    \end{equation}
	   obtained in the limit when the tree-level scalar masses  are identical. To be specific, for  $\tilde{\lambda}=0.1$ 
	   it is about $0.1\%$  of the SM value and grows to $3.5\%$ only for extremely large values of $\tilde{\lambda}=10$.
Measuring this coupling at colliders would therefore constitute an interesting probe of the model.  

\section{Properties of the elementary goldstone dark matter}
    \label{sec:DM}
    
As a candidate for DM we have the remaining SM neutral pGB, i.e. $\Pi_5$.  The stability is ensured via a discrete $Z_2$ symmetry. This is a crucial difference with respect to the composite case   \cite{Cacciapaglia:2014uja} because there the particle decays via a Wess--Zumino--Witten term induced by the underlying dynamics~\cite{Wess:1971yu,Witten:1983ar,Witten:1983tw}. In the following we will therefore investigate the associated DM phenomenology. 
 
    \subsection{Relic abundance}
    
We assume that the DM candidate is a thermal relic whose present density is determined by its annihilation processes in the early universe.     
To estimate the relic abundance, we need the couplings to the SM fields. Because $\Pi_5$ is a singlet under the electroweak gauge symmetries, it couples to the SM only via the scalar sector. Therefore, we consider the scalar mediated annihilations into scalars, EW gauge bosons and SM fermions.
	
The relevant couplings 	are
	\begin{equation}
	    \label{eq:couplings}
	    \begin{split}
		&\lambda_{\Pi_5\Pi_5hh}=\frac{M_S^2}{v^2},\quad \lambda_{\Pi_5\Pi_5h}=\frac{M_S^2\cos\alpha}{v},\quad
		    \lambda_{\Pi_5\Pi_5H}=-\frac{M_S^2\sin\alpha}{v},\\
		&\lambda_{hhh}=\frac{3M_S^2\cos\alpha}{v},\quad\lambda_{hhH}=-\frac{M_S^2\sin\alpha}{v},\quad\\
		&g_{hWW}=\frac{1}{2}g^2v\sin\theta\sin(\theta+\alpha),\quad 
		    g_{hZZ}=\frac{1}{2}(g^2+g^{\prime\, 2})v\sin\theta\sin(\theta+\alpha),\\
		&g_{HWW}=\frac{1}{2}g^2v\sin\theta\cos(\theta+\alpha),\quad 
		    g_{HZZ}=\frac{1}{2}(g^2+g^{\prime\, 2})v\sin\theta\cos(\theta+\alpha),\\
		&y_{hff}=\frac{y_f}{\sqrt{2}}\sin(\theta+\alpha),\quad y_{Hff}=\frac{y_f}{\sqrt{2}}\cos(\theta+\alpha).
	    \end{split}
	\end{equation}
All the tree-level heavy scalar masses have been set to the common value $M_S$. The number density, $n$,
	of the thermal relic can be solved  via the Lee--Weinberg equation~\cite{Lee:1977ua},
	\begin{equation}
	    \label{eq:LW}
	    \frac{\partial f(x)}{\partial x}=\frac{\langle v\sigma\rangle m_{\mathrm{DM}}^3 x^2}{H}(f^2(x)-f_{\rm{eq}}^2(x)),
	\end{equation}
	where $f(x)=n(x)/s_E$ and $x=s_E^{1/3}/m_{\mathrm{DM}}$. Here $s_E$ is the entropy density at the temperature $T$, 
	$m_{\mathrm{DM}}$ is the mass of the DM candidate and ${H}$ is the Hubble parameter.

	For the averaged cross sections, we use the integral expression~\cite{Gondolo:1990dk}
	 \begin{equation}
	    \label{eq:vsigma}
	    \langle v\sigma\rangle=\frac{1}{8m_{\mathrm{DM}}^4 T K_2^2(m_{\mathrm{DM}}/T)}\int_{4 m_{\mathrm{DM}}^2}^
		\infty \dd s\sqrt{s} (s-4 m_{\mathrm{DM}}^2)
		K_1(\sqrt{s}/T)\sigma_{\rm{tot}}(s),
	\end{equation}
	where $K_i(y)$ are the modified Bessel functions of the second kind, $s$ is the standard Mandelstam variable and $\sigma_{\rm{tot}}(s)$ is
	the total cross section for the annihilation of two DM particles to two scalars, EW gauge bosons or SM fermions computed at tree level. 
	Given the cross sections, we can now determine the present ratio $f(0)$ of the
	number to its entropy density. The fractional density parameter,
	$\Omega_{\rm{DM}}$, can then be computed via
	\begin{equation}
	    \label{eq:omegaDM}
	    \Omega_{\rm{DM}}\simeq 4.01\cdot 10^8\, m_{\mathrm{DM}} f(0).
	\end{equation}

	We define the fraction of the full amount of observed cold DM
	\begin{equation}
	    \label{eq:freldef}
	    f_{\mathrm{rel}}=\Omega_{\mathrm{DM}} {h}^2/(\Omega  {h}^2)_{\mathrm{c}},
	\end{equation}
	where $(\Omega  {h}^2)_{\mathrm{c}}=0.12$ from Planck results~\cite{Ade:2013zuv} and ${h}$ is the present value of the Hubble constant.
	
	\begin{figure}
	    \begin{center}
		\includegraphics[width=0.48\textwidth]{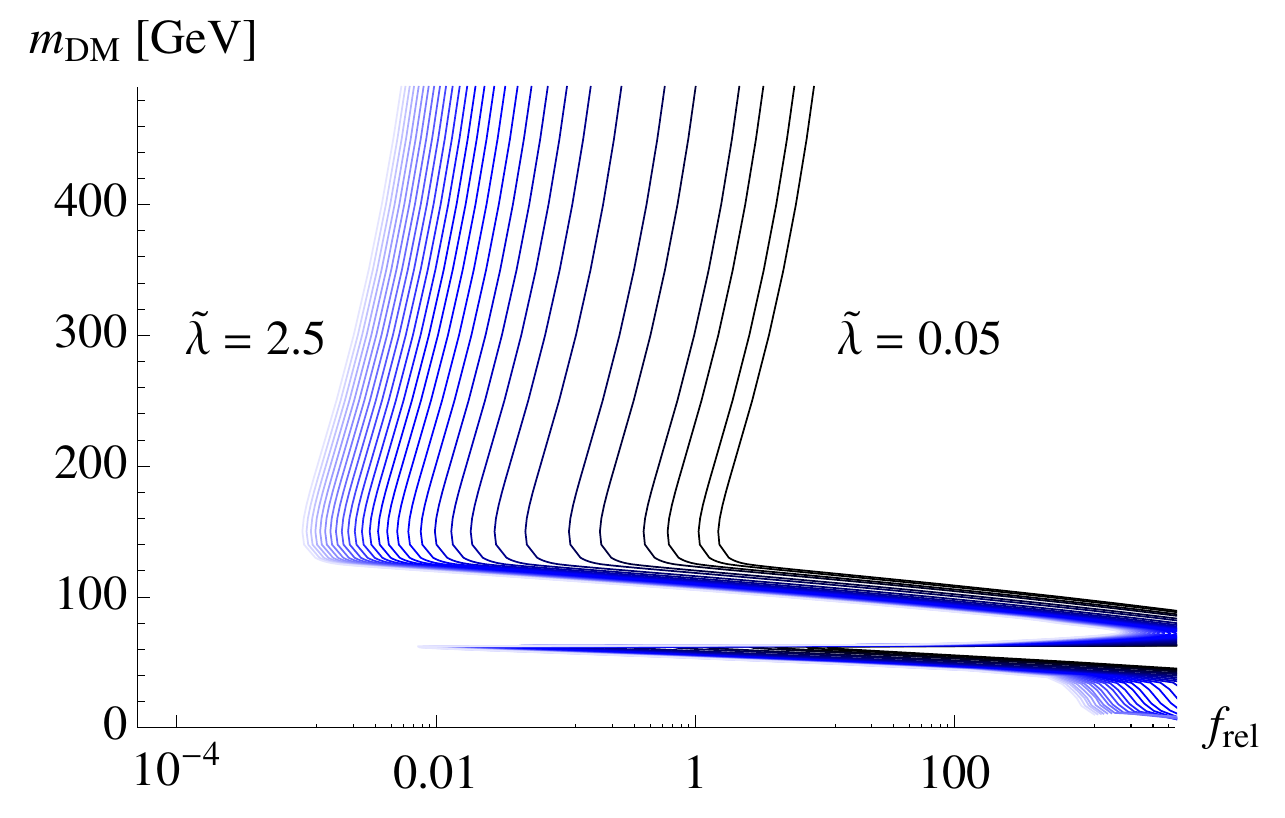}\quad
		\includegraphics[width=0.48\textwidth]{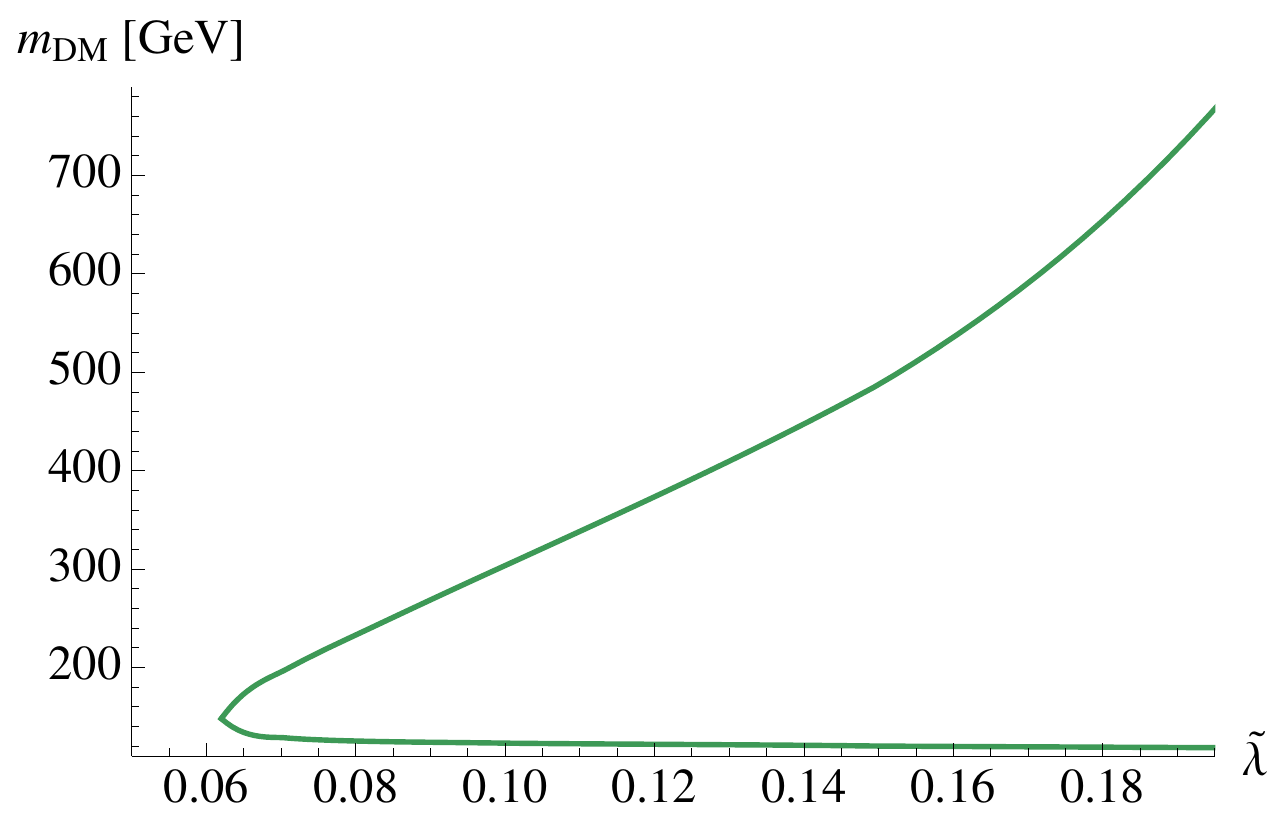}

	    \end{center}
	    \caption{{\bf Left panel:} The mass of the DM candidate as a function of the fraction of the total observed 
	    DM abundance, $f_{\mathrm{rel}}$
	    for $\tilde{\lambda}$ varying from $0.05$ to $2.5$. The curves correspond to parameter values that fulfil 
	    the minimization condition with respect to the vacuum angle, $\theta$, produce the observed EW spectrum, 
	    and give the correct Higgs mass. {\bf Right panel:} The contour giving $f_{\mathrm{rel}}=1$ in the 
	    $(\tilde{\lambda},m_{\mathrm{DM}})$ plane.}
	    \label{fig:frel}
	\end{figure}
For illustration we show $f_{\rm rel}$ in Fig.~\ref{fig:frel} as function of the DM mass for different values of $\tilde{\lambda}$ keeping fixed the phenomenologically viable spectrum for the SM. The freedom in choosing the DM mass comes from the explicit term of $\SU(4)$ symmetry breaking which provides a direct mass to $\Pi_5$. 

Moreover, we determine the DM mass as function of $\tilde{\lambda}$ once the $f_{\rm rel} = 1$ is chosen that corresponds to the observed value of  the DM relic density. The result is depicted in the right panel of Fig.~\ref{fig:frel}.

The correct relic density can be obtained either for $m_{\mathrm{DM}}\gtrsim m_h$ or $m_{\mathrm{DM}}\sim m_h/2$. 
	As we shall see, for  $m_{\mathrm{DM}}\gtrsim m_h$ the model is compatible with the experimental constraints.

    \subsection{Phenomenological constraints}
We now turn our attention to the experimental constraints coming from direct and indirect DM searches. 
	\subsubsection{Direct detection}
	  In this case the experiments directly constrain the interactions between DM and ordinary nuclei. 
	    Currently the most stringent constraints on this cross section are given by the LUX experiment \cite{Akerib:2013tjd}. 
	    In our model, the couplings $\lambda_{\Pi_5\Pi_5h/H}$ act as portals 
	    through which our DM candidate impacts on the  spin-independent scattering cross sections on nuclei. 
	    
	    The Higgs-nucleon coupling is given by $f_N m_N/v$, where $m_N=0.946$ GeV, and $f_N$ is the	    normalized total quark scalar current within the nucleon, 
	    \begin{equation}
		\label{eq:}
		f_N=\frac{1}{m_N}\sum_q\langle N|m_q\bar{q}q| N\rangle.
	    \end{equation}
	    We neglect the small differences between neutrons and protons.

	    The quark currents of the nucleon have been a subject of an intensive lattice research supplemented by efforts 
	    applying chiral perturbation theory methods and pion nucleon scattering. Consequently, $f_N$ is fairly well 
	    determined. Following~\cite{Cline:2013gha, Alanne:2014bra} we use $f_N=0.345\pm 0.016$, where the uncertainty in $f_N$ 
	    induces at most 20\% error in the spin-independent direct detection limits.

	    The spin-independent cross section for a WIMP scattering on nuclei, $\sigma_{\rm{SI}}^0$, is therefore computed by determining the $t$-channel exchange of $h^0$ and $H^0$ in the limit $t\rightarrow 0$. We allow the possibility that our WIMP candidate forms only a fraction of the total DM abundance, and this needs to be taken into account when comparing with the direct searches. The direct search constraints on 
	    $\sigma^0_{\rm{SI}}$ are given by the experimental collaborations for $f_{\rm{rel}}=1$. 
	    To apply the constraints under the assumption of subdominant WIMPs, we define an effective cross section
	    \begin{equation}
		\label{eq:}
		\sigma_{\rm{SI}}^{\rm{eff}}=f_{\rm{rel}}\sigma^0_{\rm{SI}}.
	    \end{equation}
	    
	    \begin{figure}
		\begin{center}
		    \includegraphics[width=0.6\textwidth]{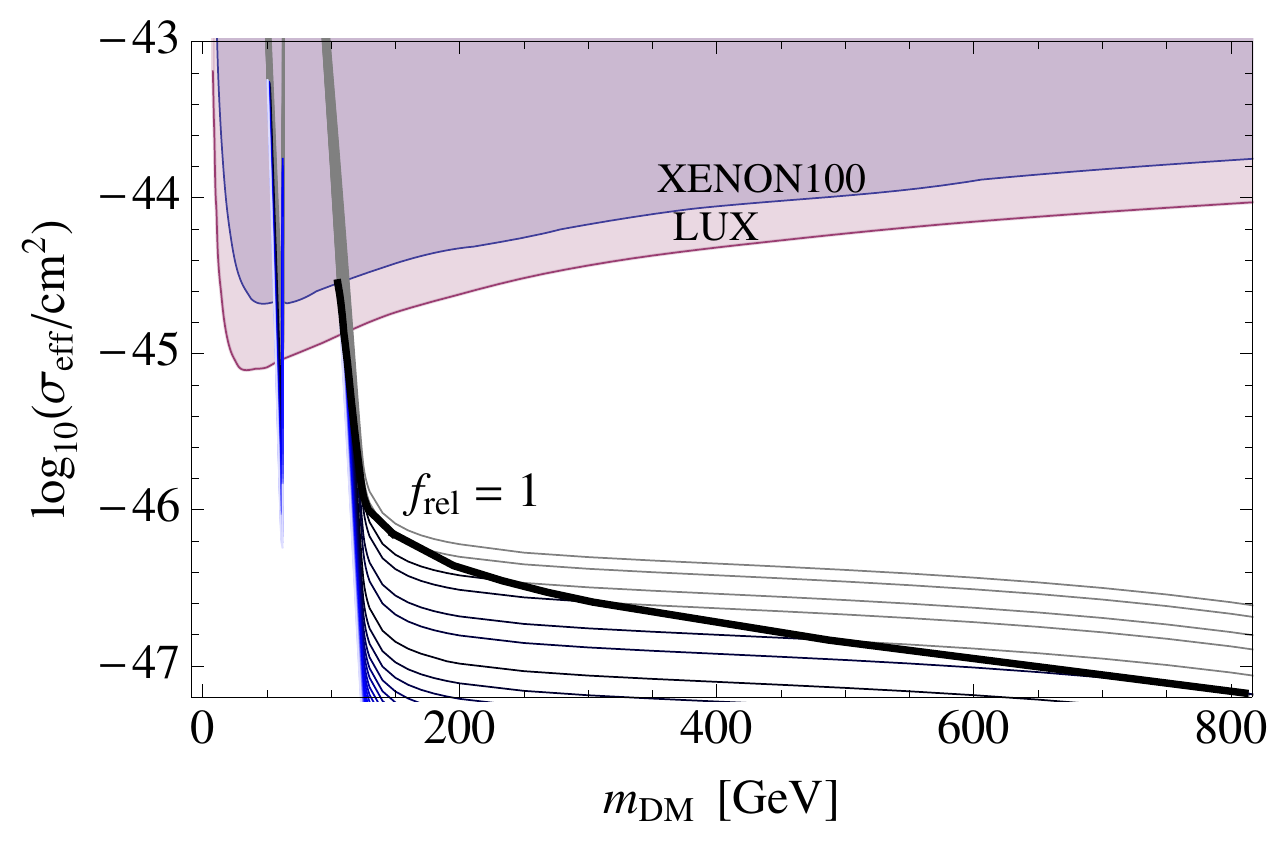}
		\end{center}
		\caption{The spin-independent cross section as a function of the mass of the DM candidate for the 
		the same cases already depicted in Fig.~\ref{fig:frel} with the approximate $f_{\mathrm{rel}}=1$ contour. The gray
		parts produce too large DM relic abundance and are, thus, excluded.}
		\label{fig:LUX}
	    \end{figure}

	    The results with the XENON100~\cite{Aprile:2011hi} and LUX  \cite{Akerib:2013tjd} limits are shown in Fig.~\ref{fig:LUX}. 
	    We plot in the same picture the approximate $f_{\mathrm{rel}}=1$ contour. It is clear from the above that elementary pGB DM in our scenario is not yet constrained by the direct detection experiments when the mass is higher than the Higgs mass. 
	
	\subsubsection{Indirect detection}

		The Fermi-LAT experiment has gathered data from the $\gamma$ ray spectrum of 25 dwarf spheroidal satellite 
		galaxies of the Milky Way~\cite{Ackermann:2013yva}. 
		These dwarf galaxies are a promising source for indirect detection of DM due to their high 
		DM content. No statistically significant excess on the $\gamma$ ray spectrum has been found
		so far, but the Fermi LAT collaboration has combined data from 15 dwarf galaxies to set an upper
		limit for the annihilation cross section of DM to different SM channels assuming the 
		Navarro--Frenk--White DM profile~\cite{Navarro:1996gj}. 
		Currently the most stringent bound for the annihilation cross section to 
		$b\bar{b}$ channel at $95\%$ C.L. for  DM particles with mass below 10 GeV is 
		$\langle\sigma v\rangle\leq3\times10^{-26}~\mathrm{cm}^3 \mathrm{s}^{-1}$.
New preliminary results seem to push this limit even further and exclude the thermal WIMP scenarios up to nearly 100~GeV~\cite{Anderson}. 

		In our model, already from the direct detection experiments along with the solution of the correct relic density we found that the most plausible region for DM was in the mass range $m_{\mathrm{DM}}\gtrsim m_h$ which is compatible with the Fermi-LAT more recent results.

\section{Conclusions}
    \label{sec:concl}
 We considered an extension of the SM featuring an enhanced global symmetry in the elementary Higgs sector yielding the $\SU(4)$ to $\Sp(4)$ pattern of spontaneous symmetry breaking.  
    
      The embedding of the electroweak gauge sector is parametrised by an angle. Its value has been fixed by minimising the quantum corrected effective potential of the theory in the presence of the electroweak and top corrections. Differently from the composite Goldstone Higgs and Technicolor scenarios, the perturbative and elementary nature of our extension enables us to precisely determine these quantum corrections. A small value of the angle is preferred by the top corrections, aligning the vacuum  where the Higgs is a pGB. The remaining pGB is neutral and becomes the DM particle. A direct comparison with  collider experiments shows that the model is phenomenologically viable. Furthermore we have shown that it is possible to obtain the observed DM thermal relic density with the model passing the most stringent experimental tests.

In the future it would be interesting to investigate the ultraviolet behaviour of the theory especially in the light of the recent  mathematical proof that certain gauge-Yukawa theories, structurally similar to the one investigated here, are ultraviolet finite because all the couplings reach a controllable interacting ultraviolet fixed point \cite{Litim:2014uca}.

\acknowledgments

We thank A. Meroni for valuable discussions. The $\mathrm{CP}^3$-Origins centre is partially funded by the Danish National Research Foundation, grant number DNRF90. 
TA would like to thank the Finnish Cultural Foundation for financial support.
KT acknowledges support from the Academy of Finland, project 267842. 

\clearpage
 
\appendix

\section{Ground state analysis and tree-level stability of the potential}
    \label{LFstability}
    We determine, at tree level, the part of parameter space of the scalar couplings for which the potential~\eqref{eq:pot} 
    remains stable at large values of the fields. For the sake of this analysis, we write the potential in terms of the scalar field components	
    \begin{equation}
	\label{eq:pot2}
	\begin{split}
	    V_M=&\frac{1}{2}m_M^2\left(\sigma^2+\Theta^2+\mathbf{\Pi}^2+\mathbf{\tilde\Pi}^2\right)
		+\frac{1}{2}c_{MR}\left(-\sigma^2+\Theta^2-\mathbf{\Pi}^2+\mathbf{\tilde\Pi}^2\right)\\
	    &+c_{MI}\left(\sigma\Theta-\mathbf{\Pi}\cdot\mathbf{\tilde\Pi}\right)
		+\frac{\lambda}{4}\left(\sigma^2+\Theta^2+\mathbf{\Pi}^2+\mathbf{\tilde\Pi}^2\right)^2\\
	    &+\lambda_1\left[\frac{1}{4}\left(\sigma^2+\Theta^2+\mathbf{\Pi}^2+\mathbf{\tilde\Pi}^2\right)^2
		+\sigma^2\mathbf{\tilde\Pi}^2+\Theta^2\mathbf{\Pi}^2+\mathbf{\Pi}^2\mathbf{\tilde\Pi}^2\right.\\
	    &\left.\qquad+2\sigma\Theta(\mathbf{\Pi}\cdot\mathbf{\tilde\Pi})-(\mathbf{\Pi}\cdot\mathbf{\tilde\Pi})^
		2\right]\\
	    &-\lambda_{2R}\left[\frac{1}{4}\left(\sigma^2-\Theta^2+\mathbf{\Pi}^2-\mathbf{\tilde\Pi}^2\right)^2
		-\left(\sigma\Theta-\mathbf{\Pi}\cdot\mathbf{\tilde\Pi}\right)^2\right]\\
	    &-\lambda_{2I}\left(\sigma^2-\Theta^2+\mathbf{\Pi}^2-\mathbf{\tilde\Pi}^2\right)
		(\mathbf{\Pi}\cdot\mathbf{\tilde\Pi}-\sigma\Theta)\\
	    &+\left[\frac{\lambda_{3R}}{4}\left(-\sigma^2+\Theta^2-\mathbf{\Pi}^2+\mathbf{\tilde\Pi}^2\right)
		+\frac{\lambda_{3I}}{2}\left(\sigma\Theta-\mathbf{\Pi}\cdot\mathbf{\tilde\Pi}\right)\right]\\
	    &\quad\cdot\left(\sigma^2+\Theta^2+\mathbf{\Pi}^2+\mathbf{\tilde\Pi}^2\right),
	\end{split}
    \end{equation}

    where $\mathbf{\Pi}^2=\Pi_i\Pi_i,\ \mathbf{\tilde\Pi}^2=\tilde\Pi_i\tilde\Pi_i$, and  
    $\mathbf{\Pi}\cdot\mathbf{\tilde\Pi}=\Pi_i\tilde\Pi_i$.
    This can be written in a more useful form by introducing the following sextuplets:
    \begin{equation}
	\label{eq:sext}
	\varphi_1=(\sigma,\ii\mathbf{\Pi}),\quad\text{and}\quad\varphi_2=(\Theta,-\ii\mathbf{\tilde{\Pi}}).
    \end{equation}
    Then 
    \begin{equation}
	\label{eq:phis}
	\varphi_1^{\dagger}\varphi_1=\sigma^2+\mathbf{\Pi}^2,\quad \varphi_2^{\dagger}\varphi_2=\Theta^2+\mathbf{\tilde{\Pi}}^2,\quad
	    \text{and}\quad\varphi_1^{\dagger}\varphi_2=\sigma\Theta-\mathbf{\Pi}\cdot\mathbf{\tilde\Pi},
    \end{equation}
    and the potential reads
    \begin{equation}
	\label{eq:potential2}
	\begin{split}
	    V_M=&\frac{1}{2}m_M^2(\varphi_1^{\dagger}\varphi_1+\varphi_2^{\dagger}\varphi_2)+\frac{1}{2}c_{MR}(-\varphi_1^{\dagger}\varphi_1
		+\varphi_2^{\dagger}\varphi_2)
		+c_{MI}\varphi_1^{\dagger}\varphi_2\\
	    &+\frac{\lambda}{4}(\varphi_1^{\dagger}\varphi_1+\varphi_2^{\dagger}\varphi_2)^2\\
	    &+\lambda_1\left[\frac{1}{4}(\varphi_1^{\dagger}\varphi_1+\varphi_2^{\dagger}\varphi_2)^2+(\varphi_1^{\dagger}\varphi_1)
		(\varphi_2^{\dagger}\varphi_2)-(\varphi_1^{\dagger}\varphi_2)^2\right]\\
	    &-\lambda_{2R}\left[\frac{1}{4}(\varphi_1^{\dagger}\varphi_1-\varphi_2^{\dagger}\varphi_2)^2
		-(\varphi_1^{\dagger}\varphi_2)^2\right]-\lambda_{2I}(-\varphi_1^{\dagger}\varphi_1+\varphi_2^{\dagger}\varphi_2)
		\varphi_1^{\dagger}\varphi_2\\
	    &\left[\frac{\lambda_{3R}}{4}(-\varphi_1^{\dagger}\varphi_1+\varphi_2^{\dagger}\varphi_2)
		+\frac{\lambda_{3I}}{2}\varphi_1^{\dagger}\varphi_2\right](\varphi_1^{\dagger}\varphi_1+\varphi_2^{\dagger}\varphi_2).
	\end{split}
    \end{equation}
    
    Limiting ourselves only to real paramters, i.e. setting $c_{MI}=\lambda_{2I}=\lambda_{3I}=0$, 
    we find the following four stationary configurations of the fields
    \begin{align}
	&\langle\varphi_1^{\dagger}\varphi_1\rangle=0,\quad  \langle\varphi_2^{\dagger}\varphi_2\rangle=0 \label{eq:max1}\\
	    ~\notag\\
	&\langle\varphi_1^{\dagger}\varphi_1\rangle=\frac{c_{MR}-m_M^2}{\lambda+\lambda_1-\lambda_{2R} -\lambda_{3R}},\quad
	    \langle\varphi_2^{\dagger}\varphi_2\rangle=0 \label{eq:min1}\\
	    ~\notag\\
	&\langle\varphi_1^{\dagger}\varphi_1\rangle=0,\quad \langle\varphi_2^{\dagger}\varphi_2\rangle
	    =\frac{-c_{MR}-m_M^2}{\lambda+\lambda_1-\lambda_{2R} +\lambda_{3R}}\label{eq:min2}\\
	    ~\notag\\
	\begin{split}
	    &\langle\varphi_1^{\dagger}\varphi_1\rangle=\frac{m_M^2(\lambda_{3R}-2(\lambda_1+\lambda_{2R} ))
		-c_{MR}(2\lambda+4\lambda_1+\lambda_{3R})}{4(\lambda+2\lambda_1)(\lambda_1+\lambda_{2R} )+\lambda_{3R}^2},\\
	    &\langle\varphi_2^{\dagger}\varphi_2\rangle=\frac{-m_M^2(\lambda_{3R}+2(\lambda_1+\lambda_{2R} ))
		+c_{MR}(2\lambda+4\lambda_1-\lambda_{3R})}{4(\lambda+2\lambda_1)(\lambda_1+\lambda_{2R} )+\lambda_{3R}^2},\\
	    &\langle\varphi_1^{\dagger}\varphi_2\rangle=0.
	\end{split}\label{eq:max2}
    \end{align}
    The second and the third are minima (if the couplings are such that the conditions give positive moduli), whereas the first and the 
    fourth are maxima. The potential at the minimum then reads
    \begin{equation}
	\label{eq:VatMin}
	V_M\left(\langle\varphi_i^{\dagger}\varphi_i\rangle\right)=-\frac{1}{4}\lambda_{\mathrm{eff},\, i}
	    \langle\varphi_i^{\dagger}\varphi_i\rangle^2,
    \end{equation}
    where $i=1,2$, no summation implied, and 
    \begin{equation}
	\label{eq:lambdas}
	\begin{split}
	    &\lambda_{\mathrm{eff},\, 1}=\lambda+\lambda_1-\lambda_{2R} -\lambda_{3R}, \\
	    &\lambda_{\mathrm{eff}\, ,2}=\lambda+\lambda_1-\lambda_{2R} +\lambda_{3R}.
	\end{split}
    \end{equation}
    The minima~\eqref{eq:min1} and~\eqref{eq:min2} are preserved also in the case of complex couplings, if the 
    imaginary parts fulfil the following conditions:
    \begin{equation}
	\label{eq:constrIm1}
	c_{MI}=-\frac{(c_{MR}-m_M^2)(2\lambda_{2I}+\lambda_{3I})}
	    {2(\lambda+\lambda_1-\lambda_{2R} -\lambda_{3R})}, 
    \end{equation}
    in case of the minimum~\eqref{eq:min1}, and
    \begin{equation}
	\label{eq:constrIm2}
	c_{MI}=-\frac{(c_{MR}+m_M^2)(2\lambda_{2I}-\lambda_{3I})}
	    {2(\lambda+\lambda_1-\lambda_{2R} +\lambda_{3R})}, 
    \end{equation}
    in case of the minimum~\eqref{eq:min2}.

    Overall stability of the potential relies on the positivity of the quartic interactions for large values of the fields. 
    This is equivalent to requiring the quartic potential to be positive definite on the unit sphere of the the scalar fields 
    manifold, i.e. $\varphi_1^{\dagger}\varphi_1+\varphi_2^{\dagger}\varphi_2=1$. Using \eqref{eq:potential2} and the constraint 
    of being on a unit sphere one derives the inequalities

    \begin{eqnarray}
	&(\varphi_1^{\dagger}\varphi_1+\varphi_2^{\dagger}\varphi_2)^2=1,\label{eq:VSterm1}\\
	&\frac{1}{4}\leq\frac{1}{4}(\varphi_1^{\dagger}\varphi_1+\varphi_2^{\dagger}\varphi_2)^2+(\varphi_1^{\dagger}\varphi_1)
		(\varphi_2^{\dagger}\varphi_2)-(\varphi_1^{\dagger}\varphi_2)^2\leq\frac{1}{2},\label{eq:VSterm2}\\
	&-\frac{1}{4}\leq\frac{1}{4}(\varphi_1^{\dagger}\varphi_1-\varphi_2^{\dagger}\varphi_2)^2
		-(\varphi_1^{\dagger}\varphi_2)^2\leq\frac{1}{4},\label{eq:VSterm3}\\
	&-\frac{1}{4}\leq(-\varphi_1^{\dagger}\varphi_1+\varphi_2^{\dagger}\varphi_2)
		\varphi_1^{\dagger}\varphi_1\leq\frac{1}{4},\label{eq:VSterm4}\\
	&0\leq(\varphi_1^{\dagger}\varphi_1-\varphi_2^{\dagger}\varphi_2)^2+4(\varphi_1^{\dagger}\varphi_2)^2
	    \leq 1,\label{eq:VSterm5}\\
	&-1\leq-\varphi_1^{\dagger}\varphi_1+\varphi_2^{\dagger}\varphi_2\leq 1,\label{eq:VSterm6}\\
	&-\frac{1}{2}\leq\varphi_1^{\dagger}\varphi_2\leq\frac{1}{2}\label{eq:VSterm7}.
    \end{eqnarray}

    We can now write a sufficient condition for the tree-level stability:
    \begin{equation}
	\label{eq:VacStab}
	\lambda+\Delta_{1}\lambda_1-|\lambda_{2R}|-|\lambda_{2I}|-|\lambda_{3R}|-|\lambda_{3I}|\geq 0,
    \end{equation}
    with  
    \begin{equation}
	\label{eq:VacStabDeltas}
	\Delta_1=\left\{\begin{array}{l}1,\quad\text{if }\lambda_1\geq0\\2,\quad\text{if }\lambda_1<0 \end{array}\right. .
    \end{equation}

\section{Generators}\label{appgenerators}
     
    We want to study the vacuum, $E$, breaking a global $\SU(4)$ symmetry to the sympletic group $\Sp(4)$.
    The unbroken generators of $\SU(4)$ belonging to the $\Sp(4)$ subgroup satisfy the relation
    \begin{equation}
	\label{eq:sp4algebra}
	S^aE+ES^{a\,\mathrm{T}}=0,\qquad a=1,\dots,10,
    \end{equation}
    and we denote the broken generators by $X^i$,\ $i=1,\dots,5$.
    
    There are two inequivalent vacua that cause the breaking $\SU(4)\rightarrow \Sp(4)$ but 
    leave the electroweak sector unbroken, namely
    \begin{equation}
	\label{eq:ABvacua}
	E_{A}=\left(\begin{array}{cc}
	    \ii\sigma_2	&   0		\\
	    0		&   \ii\sigma_2
	\end{array}\right),\quad\text{and}\quad
	E_{B}=\left(\begin{array}{cc}
	    \ii\sigma_2	&   0		\\
	    0		&   -\ii\sigma_2
	\end{array}\right).
    \end{equation}
    The vacua are inequivalent in the sense that they 
    cannot be related by an $\SU(2)_{\mathrm{L}}$ transformation. In this paper we choose to study
    the vacuum $B$.
    
    Another vacuum of interest, associated with the breaking $\SU(4)\rightarrow \Sp(4)$, is the 
    so-called technicolor vacuum that breaks the electroweak symmetry 
    ${\displaystyle \SU(2)_{\mathrm{L}}\times
    \mathrm{U}(1)_Y\rightarrow \mathrm{U}(1)_{Q}}$. This vacuum reads
    \begin{equation}
	\label{eq:Hvacuum}
	E_{H}=\left(\begin{array}{cc}
	    0   &   1	\\
	    -1	&   0
	\end{array}\right).
    \end{equation}
    
    In this paper, we want to study a linear combination of these two types of vacua: a vacuum interpolating
    between the one leaving the electroweak symmetry unbroken and the other associated with the desired 
    breaking pattern. Thus, we parameterize the mixed vacuum as
    \begin{equation}
	\label{eq:vacuum}
	E_{\theta}=\cos\theta E_B+\sin\theta E_H.
    \end{equation}
    This satisfies $E_{\theta}^{\dagger}E_{\theta}=1$.
    We give the representation of the unbroken and broken generators of $\SU(4)$ associated with $E_{\theta}$:

    \begin{equation}
	\label{eq:}
	\begin{split}
	    &S_{\theta}^{1,2,3}=
	    \frac{1}{2\sqrt{2}}
	    \left(
		\begin{array}{cc}
		\sigma_{i}   & 0 \\
		0		    & -\sigma_{i}^{T}
		\end{array}
	    \right),\quad
	    S_{\theta}^{4}=\frac{1}{2\sqrt{2}}\left(
		\begin{array}{cc}
		0   & \ii\sigma_{1} \\
		-\ii\sigma_{1}		    & 0
		\end{array}
	    \right),\quad
	    S_{\theta}^{5}=\frac{1}{2\sqrt{2}}\left(
		\begin{array}{cc}
		0   & \ii\sigma_{3} \\
		-\ii\sigma_{3}		    & 0
		\end{array}
	    \right),\\
	    &S_{\theta}^{6}=\frac{1}{2\sqrt{2}}\left(
		\begin{array}{cc}
		0   & 1 \\
		1		    & 0
		\end{array}
	    \right),\\
   	    &S_{\theta}^{7}=\frac{\cos\theta}{2\sqrt2}\left(
		\begin{array}{cc}
		\sigma_{1} &0\\
		0& \sigma_{1}^{T}
		\end{array}
	    \right)
	    +\frac{ \sin\theta  }{2\sqrt{2}}\left(
		\begin{array}{cc}
		0   & \sigma_{3} \\
		\sigma_{3}		    & 0
		\end{array}
	    \right),\quad
	    S_{\theta}^{8}=\frac{\cos\theta}{2\sqrt2}\left(
		\begin{array}{cc}
		\sigma_{2} &0\\
		0& \sigma_{2}^{T}
		\end{array}
	    \right)
	    -\frac{ \sin\theta  }{2\sqrt{2}}\left(
		\begin{array}{cc}
		0   & \ii \\
		-\ii		    & 0
		\end{array}
	    \right),\\
	    &S_{\theta}^{9}=\frac{\cos\theta}{2\sqrt2}\left(
		\begin{array}{cc}
		\sigma_{3} &0\\
		0& \sigma_{3}^{T}
		\end{array}
	    \right)
	    -\frac{ \sin\theta  }{2\sqrt{2}}\left(
		\begin{array}{cc}
		0   & \sigma_{1} \\
		\sigma_{1}		    & 0
		\end{array}
	    \right),\quad
	    S_{\theta}^{10}=\frac{\cos\theta}{2\sqrt2}\left(
		\begin{array}{cc}
		0 & \ii\sigma_{2}\\
		-\ii\sigma_{2} & 0
		\end{array}
	    \right)
	    +\frac{ \sin\theta  }{2\sqrt{2}}\left(
		\begin{array}{cc}
		1   & 0 \\
		0    & -1
		\end{array}
	    \right).
	\end{split}
    \end{equation}
    and
    \begin{equation}
	\begin{split}
	    \label{eq:}
	    &X_{\theta}^{1}=\frac{\cos\theta}{2\sqrt{2}} \left(
		\begin{array}{cc}
		0   &  \sigma_{3}\\
		\sigma_{3}    & 0
		\end{array}
	    \right)-\frac{\sin\theta}{2\sqrt{2}}\left(
		\begin{array}{cc}
		\sigma_{1}   & 0 \\
		0    & \sigma_{1}^{T}
		\end{array}
	    \right),\quad
	    X_{\theta}^{2}=\frac{\cos\theta}{2\sqrt{2}} \left(
		\begin{array}{cc}
		0   &  \ii\\
		-\ii    & 0
		\end{array}
	    \right)+\frac{\sin\theta}{2\sqrt{2}}\left(
		\begin{array}{cc}
		\sigma_{2}   & 0 \\
		0    & \sigma_{2}^{T}
		\end{array}
	    \right)\\
	    &X_{\theta}^{3}=\frac{\cos\theta}{2\sqrt{2}} \left(
		\begin{array}{cc}
		0   &  \sigma_{1}\\
		\sigma_{1}    & 0
		\end{array}
	    \right)+\frac{\sin\theta}{2\sqrt{2}}\left(
		\begin{array}{cc}
		\sigma_{3}   & 0 \\
		0    & \sigma_{3}^{T}
		\end{array}
	    \right),\quad
	    X_{\theta}^{4}=\frac{1}{2\sqrt{2}} \left(
		\begin{array}{cc}
		0   &  \sigma_{2}\\
		\sigma_{2}    & 0
		\end{array}
	    \right),\\
   	    &X_{\theta}^{5}=\frac{\cos\theta}{2\sqrt{2}} \left(
		\begin{array}{cc}
		1  &  0\\
		0    & -1
		\end{array}
	    \right)-\frac{\sin\theta}{2\sqrt{2}}\left(
		\begin{array}{cc}
		0 & \ii\sigma_{2}   \\
		-\ii \sigma_{2} & 0
		\end{array}
	    \right).
	\end{split}
    \end{equation}

    The  generators are normalized as 
    \begin{equation}
	\label{eq:genNorm}
	\mathrm{Tr}[S_{\theta}^aS_{\theta}^b]=\frac{1}{2}\delta^{ab},\quad \mathrm{Tr}[X_{\theta}^iX_{\theta}^j]
	    =\frac{1}{2}\delta^{ij}.
    \end{equation}

\section{Pfaffians and determinants}\label{pfaffians}
    Pfaffian of a $2n\times 2n$ antisymmetric matrix is given by
    \begin{equation}
	\label{eq:Pf}
	\Pf (A)=\frac{1}{2^n n!}\sum_{\sigma\in \mathrm{S}_{2n}}\mathrm{sgn}(\sigma)
	    \prod_i A_{\sigma(2i-1)\sigma(2i)},
    \end{equation}
    where $\mathrm{S}_{2n}$ is the group of permutations of $2n$ elements and $\mathrm{sgn}(\sigma)$ is the
    sign of the permutation $\sigma$.

    The Pfaffian of a $4\times 4$ antisymmetric matrix $A$ reads then
    \begin{equation}
	\label{eq:PfA}
	\Pf(A)=\frac{1}{8}\epsilon_{ijkl}A_{ij}A_{kl},
    \end{equation}
    where $\epsilon_{ijkl}$ is the Levi-Civita symbol, $\epsilon_{1234}=1$.

    The determinant of an antisymmetric matrix $A$ equals the Pfaffian of the matrix squared, i.e.
    $\Pf(A)^2=\mathrm{Det}(A).$ Explicitly, for
    $4\times 4$ antisymmetric matrix, $A$, this equality can be written as
    \begin{equation}
	\label{eq:PfDet}
	\begin{split}
	\frac{1}{8^2}\epsilon_{i_1i_2i_3i_4}\epsilon_{j_1j_2j_3j_4}
	    A_{i_1i_2}A_{i_3i_4}A_{j_1j_2}A_{j_3j_4}
	  =\frac{1}{4!}\epsilon_{i_1i_2i_3i_4}\epsilon_{j_1j_2j_3j_4}
	    A_{i_1j_1}A_{i_2j_2}A_{i_3j_3}A_{i_4j_4}.
	\end{split}
    \end{equation}

%%%%%%%%%%%%%%%%%%%%%%%%%%%%%%%%%%%%    
\bibliography{Elementary}

\begin{thebibliography}{40}%
\makeatletter
\providecommand \@ifxundefined [1]{%
 \@ifx{#1\undefined}
}%
\providecommand \@ifnum [1]{%
 \ifnum #1\expandafter \@firstoftwo
 \else \expandafter \@secondoftwo
 \fi
}%
\providecommand \@ifx [1]{%
 \ifx #1\expandafter \@firstoftwo
 \else \expandafter \@secondoftwo
 \fi
}%
\providecommand \natexlab [1]{#1}%
\providecommand \enquote  [1]{``#1''}%
\providecommand \bibnamefont  [1]{#1}%
\providecommand \bibfnamefont [1]{#1}%
\providecommand \citenamefont [1]{#1}%
\providecommand \href@noop [0]{\@secondoftwo}%
\providecommand \href [0]{\begingroup \@sanitize@url \@href}%
\providecommand \@href[1]{\@@startlink{#1}\@@href}%
\providecommand \@@href[1]{\endgroup#1\@@endlink}%
\providecommand \@sanitize@url [0]{\catcode `\\12\catcode `\$12\catcode
  `\&12\catcode `\#12\catcode `\^12\catcode `\_12\catcode `\%12\relax}%
\providecommand \@@startlink[1]{}%
\providecommand \@@endlink[0]{}%
\providecommand \url  [0]{\begingroup\@sanitize@url \@url }%
\providecommand \@url [1]{\endgroup\@href {#1}{\urlprefix }}%
\providecommand \urlprefix  [0]{URL }%
\providecommand \Eprint [0]{\href }%
\providecommand \doibase [0]{http://dx.doi.org/}%
\providecommand \selectlanguage [0]{\@gobble}%
\providecommand \bibinfo  [0]{\@secondoftwo}%
\providecommand \bibfield  [0]{\@secondoftwo}%
\providecommand \translation [1]{[#1]}%
\providecommand \BibitemOpen [0]{}%
\providecommand \bibitemStop [0]{}%
\providecommand \bibitemNoStop [0]{.\EOS\space}%
\providecommand \EOS [0]{\spacefactor3000\relax}%
\providecommand \BibitemShut  [1]{\csname bibitem#1\endcsname}%
\let\auto@bib@innerbib\@empty
%</preamble>
\bibitem [{\citenamefont {Sannino}\ and\ \citenamefont
  {Schechter}(1999)}]{Sannino:1999qe}%
  \BibitemOpen
  \bibfield  {author} {\bibinfo {author} {\bibfnamefont {F.}~\bibnamefont
  {Sannino}}\ and\ \bibinfo {author} {\bibfnamefont {J.}~\bibnamefont
  {Schechter}},\ }\href {\doibase 10.1103/PhysRevD.60.056004} {\bibfield
  {journal} {\bibinfo  {journal} {Phys.Rev.}\ }\textbf {\bibinfo {volume}
  {D60}},\ \bibinfo {pages} {056004} (\bibinfo {year} {1999})},\ \Eprint
  {http://arxiv.org/abs/hep-ph/9903359} {arXiv:hep-ph/9903359 [hep-ph]}
  \BibitemShut {NoStop}%
%%CITATION = HEP-PH/9903359;%%
\bibitem [{\citenamefont {Sannino}\ and\ \citenamefont
  {Tuominen}(2005)}]{Sannino:2004qp}%
  \BibitemOpen
  \bibfield  {author} {\bibinfo {author} {\bibfnamefont {F.}~\bibnamefont
  {Sannino}}\ and\ \bibinfo {author} {\bibfnamefont {K.}~\bibnamefont
  {Tuominen}},\ }\href {\doibase 10.1103/PhysRevD.71.051901} {\bibfield
  {journal} {\bibinfo  {journal} {Phys.Rev.}\ }\textbf {\bibinfo {volume}
  {D71}},\ \bibinfo {pages} {051901} (\bibinfo {year} {2005})},\ \Eprint
  {http://arxiv.org/abs/hep-ph/0405209} {arXiv:hep-ph/0405209 [hep-ph]}
  \BibitemShut {NoStop}%
%%CITATION = HEP-PH/0405209;%%
\bibitem [{\citenamefont {Dietrich}\ and\ \citenamefont
  {Sannino}(2007)}]{Dietrich:2006cm}%
  \BibitemOpen
  \bibfield  {author} {\bibinfo {author} {\bibfnamefont {D.~D.}\ \bibnamefont
  {Dietrich}}\ and\ \bibinfo {author} {\bibfnamefont {F.}~\bibnamefont
  {Sannino}},\ }\href {\doibase 10.1103/PhysRevD.75.085018} {\bibfield
  {journal} {\bibinfo  {journal} {Phys.Rev.}\ }\textbf {\bibinfo {volume}
  {D75}},\ \bibinfo {pages} {085018} (\bibinfo {year} {2007})},\ \Eprint
  {http://arxiv.org/abs/hep-ph/0611341} {arXiv:hep-ph/0611341 [hep-ph]}
  \BibitemShut {NoStop}%
%%CITATION = HEP-PH/0611341;%%
\bibitem [{\citenamefont {Foadi}\ \emph {et~al.}(2013)\citenamefont {Foadi},
  \citenamefont {Frandsen},\ and\ \citenamefont {Sannino}}]{Foadi:2012bb}%
  \BibitemOpen
  \bibfield  {author} {\bibinfo {author} {\bibfnamefont {R.}~\bibnamefont
  {Foadi}}, \bibinfo {author} {\bibfnamefont {M.~T.}\ \bibnamefont {Frandsen}},
  \ and\ \bibinfo {author} {\bibfnamefont {F.}~\bibnamefont {Sannino}},\ }\href
  {\doibase 10.1103/PhysRevD.87.095001} {\bibfield  {journal} {\bibinfo
  {journal} {Phys.Rev.}\ }\textbf {\bibinfo {volume} {D87}},\ \bibinfo {pages}
  {095001} (\bibinfo {year} {2013})},\ \Eprint {http://arxiv.org/abs/1211.1083}
  {arXiv:1211.1083 [hep-ph]} \BibitemShut {NoStop}%
%%CITATION = ARXIV:1211.1083;%%
\bibitem [{\citenamefont {Fodor}\ \emph {et~al.}(2014)\citenamefont {Fodor},
  \citenamefont {Holland}, \citenamefont {Kuti}, \citenamefont {Nogradi},\ and\
  \citenamefont {Wong}}]{Fodor:2014pqa}%
  \BibitemOpen
  \bibfield  {author} {\bibinfo {author} {\bibfnamefont {Z.}~\bibnamefont
  {Fodor}}, \bibinfo {author} {\bibfnamefont {K.}~\bibnamefont {Holland}},
  \bibinfo {author} {\bibfnamefont {J.}~\bibnamefont {Kuti}}, \bibinfo {author}
  {\bibfnamefont {D.}~\bibnamefont {Nogradi}}, \ and\ \bibinfo {author}
  {\bibfnamefont {C.~H.}\ \bibnamefont {Wong}},\ }\href@noop {} {\bibfield
  {journal} {\bibinfo  {journal} {PoS}\ }\textbf {\bibinfo {volume}
  {LATTICE2013}},\ \bibinfo {pages} {062} (\bibinfo {year} {2014})},\ \Eprint
  {http://arxiv.org/abs/1401.2176} {arXiv:1401.2176 [hep-lat]} \BibitemShut
  {NoStop}%
%%CITATION = ARXIV:1401.2176;%%
\bibitem [{\citenamefont {Grinstein}\ and\ \citenamefont
  {Uttayarat}(2011)}]{Grinstein:2011dq}%
  \BibitemOpen
  \bibfield  {author} {\bibinfo {author} {\bibfnamefont {B.}~\bibnamefont
  {Grinstein}}\ and\ \bibinfo {author} {\bibfnamefont {P.}~\bibnamefont
  {Uttayarat}},\ }\href {\doibase 10.1007/JHEP07(2011)038} {\bibfield
  {journal} {\bibinfo  {journal} {JHEP}\ }\textbf {\bibinfo {volume} {1107}},\
  \bibinfo {pages} {038} (\bibinfo {year} {2011})},\ \Eprint
  {http://arxiv.org/abs/1105.2370} {arXiv:1105.2370 [hep-ph]} \BibitemShut
  {NoStop}%
%%CITATION = ARXIV:1105.2370;%%
\bibitem [{\citenamefont {Antipin}\ \emph
  {et~al.}(2012{\natexlab{a}})\citenamefont {Antipin}, \citenamefont {Mojaza},\
  and\ \citenamefont {Sannino}}]{Antipin:2011aa}%
  \BibitemOpen
  \bibfield  {author} {\bibinfo {author} {\bibfnamefont {O.}~\bibnamefont
  {Antipin}}, \bibinfo {author} {\bibfnamefont {M.}~\bibnamefont {Mojaza}}, \
  and\ \bibinfo {author} {\bibfnamefont {F.}~\bibnamefont {Sannino}},\ }\href
  {\doibase 10.1016/j.physletb.2012.04.050} {\bibfield  {journal} {\bibinfo
  {journal} {Phys.Lett.}\ }\textbf {\bibinfo {volume} {B712}},\ \bibinfo
  {pages} {119} (\bibinfo {year} {2012}{\natexlab{a}})},\ \Eprint
  {http://arxiv.org/abs/1107.2932} {arXiv:1107.2932 [hep-ph]} \BibitemShut
  {NoStop}%
%%CITATION = ARXIV:1107.2932;%%
\bibitem [{\citenamefont {Antipin}\ \emph
  {et~al.}(2012{\natexlab{b}})\citenamefont {Antipin}, \citenamefont
  {Di~Chiara}, \citenamefont {Mojaza}, \citenamefont {M¿lgaard},\ and\
  \citenamefont {Sannino}}]{Antipin:2012kc}%
  \BibitemOpen
  \bibfield  {author} {\bibinfo {author} {\bibfnamefont {O.}~\bibnamefont
  {Antipin}}, \bibinfo {author} {\bibfnamefont {S.}~\bibnamefont {Di~Chiara}},
  \bibinfo {author} {\bibfnamefont {M.}~\bibnamefont {Mojaza}}, \bibinfo
  {author} {\bibfnamefont {E.}~\bibnamefont {M¿lgaard}}, \ and\ \bibinfo
  {author} {\bibfnamefont {F.}~\bibnamefont {Sannino}},\ }\href {\doibase
  10.1103/PhysRevD.86.085009} {\bibfield  {journal} {\bibinfo  {journal}
  {Phys.Rev.}\ }\textbf {\bibinfo {volume} {D86}},\ \bibinfo {pages} {085009}
  (\bibinfo {year} {2012}{\natexlab{b}})},\ \Eprint
  {http://arxiv.org/abs/1205.6157} {arXiv:1205.6157 [hep-ph]} \BibitemShut
  {NoStop}%
%%CITATION = ARXIV:1205.6157;%%
\bibitem [{\citenamefont {Antipin}\ \emph {et~al.}(2013)\citenamefont
  {Antipin}, \citenamefont {Mojaza},\ and\ \citenamefont
  {Sannino}}]{Antipin:2012sm}%
  \BibitemOpen
  \bibfield  {author} {\bibinfo {author} {\bibfnamefont {O.}~\bibnamefont
  {Antipin}}, \bibinfo {author} {\bibfnamefont {M.}~\bibnamefont {Mojaza}}, \
  and\ \bibinfo {author} {\bibfnamefont {F.}~\bibnamefont {Sannino}},\ }\href
  {\doibase 10.1103/PhysRevD.87.096005} {\bibfield  {journal} {\bibinfo
  {journal} {Phys.Rev.}\ }\textbf {\bibinfo {volume} {D87}},\ \bibinfo {pages}
  {096005} (\bibinfo {year} {2013})},\ \Eprint {http://arxiv.org/abs/1208.0987}
  {arXiv:1208.0987 [hep-ph]} \BibitemShut {NoStop}%
%%CITATION = ARXIV:1208.0987;%%
\bibitem [{\citenamefont {Antipin}\ \emph {et~al.}(2014)\citenamefont
  {Antipin}, \citenamefont {Mojaza},\ and\ \citenamefont
  {Sannino}}]{Antipin:2013exa}%
  \BibitemOpen
  \bibfield  {author} {\bibinfo {author} {\bibfnamefont {O.}~\bibnamefont
  {Antipin}}, \bibinfo {author} {\bibfnamefont {M.}~\bibnamefont {Mojaza}}, \
  and\ \bibinfo {author} {\bibfnamefont {F.}~\bibnamefont {Sannino}},\ }\href
  {\doibase 10.1103/PhysRevD.89.085015} {\bibfield  {journal} {\bibinfo
  {journal} {Phys.Rev.}\ }\textbf {\bibinfo {volume} {D89}},\ \bibinfo {pages}
  {085015} (\bibinfo {year} {2014})},\ \Eprint {http://arxiv.org/abs/1310.0957}
  {arXiv:1310.0957 [hep-ph]} \BibitemShut {NoStop}%
%%CITATION = ARXIV:1310.0957;%%
\bibitem [{\citenamefont {Kaplan}\ and\ \citenamefont
  {Georgi}(1984)}]{Kaplan:1983fs}%
  \BibitemOpen
  \bibfield  {author} {\bibinfo {author} {\bibfnamefont {D.~B.}\ \bibnamefont
  {Kaplan}}\ and\ \bibinfo {author} {\bibfnamefont {H.}~\bibnamefont
  {Georgi}},\ }\href {\doibase 10.1016/0370-2693(84)91177-8} {\bibfield
  {journal} {\bibinfo  {journal} {Phys.Lett.}\ }\textbf {\bibinfo {volume}
  {B136}},\ \bibinfo {pages} {183} (\bibinfo {year} {1984})}\BibitemShut
  {NoStop}%
%%CITATION = PHLTA,B136,183;%%
\bibitem [{\citenamefont {Kaplan}\ \emph {et~al.}(1984)\citenamefont {Kaplan},
  \citenamefont {Georgi},\ and\ \citenamefont {Dimopoulos}}]{Kaplan:1983sm}%
  \BibitemOpen
  \bibfield  {author} {\bibinfo {author} {\bibfnamefont {D.~B.}\ \bibnamefont
  {Kaplan}}, \bibinfo {author} {\bibfnamefont {H.}~\bibnamefont {Georgi}}, \
  and\ \bibinfo {author} {\bibfnamefont {S.}~\bibnamefont {Dimopoulos}},\
  }\href {\doibase 10.1016/0370-2693(84)91178-X} {\bibfield  {journal}
  {\bibinfo  {journal} {Phys.Lett.}\ }\textbf {\bibinfo {volume} {B136}},\
  \bibinfo {pages} {187} (\bibinfo {year} {1984})}\BibitemShut {NoStop}%
%%CITATION = PHLTA,B136,187;%%
\bibitem [{\citenamefont {Cacciapaglia}\ and\ \citenamefont
  {Sannino}(2014)}]{Cacciapaglia:2014uja}%
  \BibitemOpen
  \bibfield  {author} {\bibinfo {author} {\bibfnamefont {G.}~\bibnamefont
  {Cacciapaglia}}\ and\ \bibinfo {author} {\bibfnamefont {F.}~\bibnamefont
  {Sannino}},\ }\href {\doibase 10.1007/JHEP04(2014)111} {\bibfield  {journal}
  {\bibinfo  {journal} {JHEP}\ }\textbf {\bibinfo {volume} {1404}},\ \bibinfo
  {pages} {111} (\bibinfo {year} {2014})},\ \Eprint
  {http://arxiv.org/abs/1402.0233} {arXiv:1402.0233 [hep-ph]} \BibitemShut
  {NoStop}%
%%CITATION = ARXIV:1402.0233;%%
\bibitem [{\citenamefont {Lewis}\ \emph {et~al.}(2012)\citenamefont {Lewis},
  \citenamefont {Pica},\ and\ \citenamefont {Sannino}}]{Lewis:2011zb}%
  \BibitemOpen
  \bibfield  {author} {\bibinfo {author} {\bibfnamefont {R.}~\bibnamefont
  {Lewis}}, \bibinfo {author} {\bibfnamefont {C.}~\bibnamefont {Pica}}, \ and\
  \bibinfo {author} {\bibfnamefont {F.}~\bibnamefont {Sannino}},\ }\href
  {\doibase 10.1103/PhysRevD.85.014504} {\bibfield  {journal} {\bibinfo
  {journal} {Phys.Rev.}\ }\textbf {\bibinfo {volume} {D85}},\ \bibinfo {pages}
  {014504} (\bibinfo {year} {2012})},\ \Eprint {http://arxiv.org/abs/1109.3513}
  {arXiv:1109.3513 [hep-ph]} \BibitemShut {NoStop}%
%%CITATION = ARXIV:1109.3513;%%
\bibitem [{\citenamefont {Hietanen}\ \emph {et~al.}(2013)\citenamefont
  {Hietanen}, \citenamefont {Lewis}, \citenamefont {Pica},\ and\ \citenamefont
  {Sannino}}]{Hietanen:2013fya}%
  \BibitemOpen
  \bibfield  {author} {\bibinfo {author} {\bibfnamefont {A.}~\bibnamefont
  {Hietanen}}, \bibinfo {author} {\bibfnamefont {R.}~\bibnamefont {Lewis}},
  \bibinfo {author} {\bibfnamefont {C.}~\bibnamefont {Pica}}, \ and\ \bibinfo
  {author} {\bibfnamefont {F.}~\bibnamefont {Sannino}},\ }\href@noop {} {\
  (\bibinfo {year} {2013})},\ \Eprint {http://arxiv.org/abs/1308.4130}
  {arXiv:1308.4130 [hep-ph]} \BibitemShut {NoStop}%
%%CITATION = ARXIV:1308.4130;%%
\bibitem [{\citenamefont {Hietanen}\ \emph {et~al.}(2014)\citenamefont
  {Hietanen}, \citenamefont {Lewis}, \citenamefont {Pica},\ and\ \citenamefont
  {Sannino}}]{Hietanen:2014xca}%
  \BibitemOpen
  \bibfield  {author} {\bibinfo {author} {\bibfnamefont {A.}~\bibnamefont
  {Hietanen}}, \bibinfo {author} {\bibfnamefont {R.}~\bibnamefont {Lewis}},
  \bibinfo {author} {\bibfnamefont {C.}~\bibnamefont {Pica}}, \ and\ \bibinfo
  {author} {\bibfnamefont {F.}~\bibnamefont {Sannino}},\ }\href {\doibase
  10.1007/JHEP07(2014)116} {\bibfield  {journal} {\bibinfo  {journal} {JHEP}\
  }\textbf {\bibinfo {volume} {1407}},\ \bibinfo {pages} {116} (\bibinfo {year}
  {2014})},\ \Eprint {http://arxiv.org/abs/1404.2794} {arXiv:1404.2794
  [hep-lat]} \BibitemShut {NoStop}%
%%CITATION = ARXIV:1404.2794;%%
\bibitem [{\citenamefont {Appelquist}\ \emph {et~al.}(1999)\citenamefont
  {Appelquist}, \citenamefont {Rodrigues~da Silva},\ and\ \citenamefont
  {Sannino}}]{Appelquist:1999dq}%
  \BibitemOpen
  \bibfield  {author} {\bibinfo {author} {\bibfnamefont {T.}~\bibnamefont
  {Appelquist}}, \bibinfo {author} {\bibfnamefont {P.}~\bibnamefont
  {Rodrigues~da Silva}}, \ and\ \bibinfo {author} {\bibfnamefont
  {F.}~\bibnamefont {Sannino}},\ }\href {\doibase 10.1103/PhysRevD.60.116007}
  {\bibfield  {journal} {\bibinfo  {journal} {Phys.Rev.}\ }\textbf {\bibinfo
  {volume} {D60}},\ \bibinfo {pages} {116007} (\bibinfo {year} {1999})},\
  \Eprint {http://arxiv.org/abs/hep-ph/9906555} {arXiv:hep-ph/9906555 [hep-ph]}
  \BibitemShut {NoStop}%
%%CITATION = HEP-PH/9906555;%%
\bibitem [{\citenamefont {Duan}\ \emph {et~al.}(2001)\citenamefont {Duan},
  \citenamefont {Rodrigues~da Silva},\ and\ \citenamefont
  {Sannino}}]{Duan:2000dy}%
  \BibitemOpen
  \bibfield  {author} {\bibinfo {author} {\bibfnamefont {Z.-y.}\ \bibnamefont
  {Duan}}, \bibinfo {author} {\bibfnamefont {P.}~\bibnamefont {Rodrigues~da
  Silva}}, \ and\ \bibinfo {author} {\bibfnamefont {F.}~\bibnamefont
  {Sannino}},\ }\href {\doibase 10.1016/S0550-3213(00)00550-2} {\bibfield
  {journal} {\bibinfo  {journal} {Nucl.Phys.}\ }\textbf {\bibinfo {volume}
  {B592}},\ \bibinfo {pages} {371} (\bibinfo {year} {2001})},\ \Eprint
  {http://arxiv.org/abs/hep-ph/0001303} {arXiv:hep-ph/0001303 [hep-ph]}
  \BibitemShut {NoStop}%
%%CITATION = HEP-PH/0001303;%%
\bibitem [{\citenamefont {Ryttov}\ and\ \citenamefont
  {Sannino}(2008)}]{Ryttov:2008xe}%
  \BibitemOpen
  \bibfield  {author} {\bibinfo {author} {\bibfnamefont {T.~A.}\ \bibnamefont
  {Ryttov}}\ and\ \bibinfo {author} {\bibfnamefont {F.}~\bibnamefont
  {Sannino}},\ }\href {\doibase 10.1103/PhysRevD.78.115010} {\bibfield
  {journal} {\bibinfo  {journal} {Phys.Rev.}\ }\textbf {\bibinfo {volume}
  {D78}},\ \bibinfo {pages} {115010} (\bibinfo {year} {2008})},\ \Eprint
  {http://arxiv.org/abs/0809.0713} {arXiv:0809.0713 [hep-ph]} \BibitemShut
  {NoStop}%
%%CITATION = ARXIV:0809.0713;%%
\bibitem [{\citenamefont {Katz}\ \emph {et~al.}(2005)\citenamefont {Katz},
  \citenamefont {Nelson},\ and\ \citenamefont {Walker}}]{Katz:2005au}%
  \BibitemOpen
  \bibfield  {author} {\bibinfo {author} {\bibfnamefont {E.}~\bibnamefont
  {Katz}}, \bibinfo {author} {\bibfnamefont {A.~E.}\ \bibnamefont {Nelson}}, \
  and\ \bibinfo {author} {\bibfnamefont {D.~G.}\ \bibnamefont {Walker}},\
  }\href {\doibase 10.1088/1126-6708/2005/08/074} {\bibfield  {journal}
  {\bibinfo  {journal} {JHEP}\ }\textbf {\bibinfo {volume} {0508}},\ \bibinfo
  {pages} {074} (\bibinfo {year} {2005})},\ \Eprint
  {http://arxiv.org/abs/hep-ph/0504252} {arXiv:hep-ph/0504252 [hep-ph]}
  \BibitemShut {NoStop}%
%%CITATION = HEP-PH/0504252;%%
\bibitem [{\citenamefont {Gripaios}\ \emph {et~al.}(2009)\citenamefont
  {Gripaios}, \citenamefont {Pomarol}, \citenamefont {Riva},\ and\
  \citenamefont {Serra}}]{Gripaios:2009pe}%
  \BibitemOpen
  \bibfield  {author} {\bibinfo {author} {\bibfnamefont {B.}~\bibnamefont
  {Gripaios}}, \bibinfo {author} {\bibfnamefont {A.}~\bibnamefont {Pomarol}},
  \bibinfo {author} {\bibfnamefont {F.}~\bibnamefont {Riva}}, \ and\ \bibinfo
  {author} {\bibfnamefont {J.}~\bibnamefont {Serra}},\ }\href {\doibase
  10.1088/1126-6708/2009/04/070} {\bibfield  {journal} {\bibinfo  {journal}
  {JHEP}\ }\textbf {\bibinfo {volume} {0904}},\ \bibinfo {pages} {070}
  (\bibinfo {year} {2009})},\ \Eprint {http://arxiv.org/abs/0902.1483}
  {arXiv:0902.1483 [hep-ph]} \BibitemShut {NoStop}%
%%CITATION = ARXIV:0902.1483;%%
\bibitem [{\citenamefont {Galloway}\ \emph {et~al.}(2010)\citenamefont
  {Galloway}, \citenamefont {Evans}, \citenamefont {Luty},\ and\ \citenamefont
  {Tacchi}}]{Galloway:2010bp}%
  \BibitemOpen
  \bibfield  {author} {\bibinfo {author} {\bibfnamefont {J.}~\bibnamefont
  {Galloway}}, \bibinfo {author} {\bibfnamefont {J.~A.}\ \bibnamefont {Evans}},
  \bibinfo {author} {\bibfnamefont {M.~A.}\ \bibnamefont {Luty}}, \ and\
  \bibinfo {author} {\bibfnamefont {R.~A.}\ \bibnamefont {Tacchi}},\ }\href
  {\doibase 10.1007/JHEP10(2010)086} {\bibfield  {journal} {\bibinfo  {journal}
  {JHEP}\ }\textbf {\bibinfo {volume} {1010}},\ \bibinfo {pages} {086}
  (\bibinfo {year} {2010})},\ \Eprint {http://arxiv.org/abs/1001.1361}
  {arXiv:1001.1361 [hep-ph]} \BibitemShut {NoStop}%
%%CITATION = ARXIV:1001.1361;%%
\bibitem [{\citenamefont {Barnard}\ \emph {et~al.}(2014)\citenamefont
  {Barnard}, \citenamefont {Gherghetta},\ and\ \citenamefont
  {Ray}}]{Barnard:2013zea}%
  \BibitemOpen
  \bibfield  {author} {\bibinfo {author} {\bibfnamefont {J.}~\bibnamefont
  {Barnard}}, \bibinfo {author} {\bibfnamefont {T.}~\bibnamefont {Gherghetta}},
  \ and\ \bibinfo {author} {\bibfnamefont {T.~S.}\ \bibnamefont {Ray}},\ }\href
  {\doibase 10.1007/JHEP02(2014)002} {\bibfield  {journal} {\bibinfo  {journal}
  {JHEP}\ }\textbf {\bibinfo {volume} {1402}},\ \bibinfo {pages} {002}
  (\bibinfo {year} {2014})},\ \Eprint {http://arxiv.org/abs/1311.6562}
  {arXiv:1311.6562 [hep-ph]} \BibitemShut {NoStop}%
%%CITATION = ARXIV:1311.6562;%%
\bibitem [{\citenamefont {Ferretti}\ and\ \citenamefont
  {Karateev}(2014)}]{Ferretti:2013kya}%
  \BibitemOpen
  \bibfield  {author} {\bibinfo {author} {\bibfnamefont {G.}~\bibnamefont
  {Ferretti}}\ and\ \bibinfo {author} {\bibfnamefont {D.}~\bibnamefont
  {Karateev}},\ }\href {\doibase 10.1007/JHEP03(2014)077} {\bibfield  {journal}
  {\bibinfo  {journal} {JHEP}\ }\textbf {\bibinfo {volume} {1403}},\ \bibinfo
  {pages} {077} (\bibinfo {year} {2014})},\ \Eprint
  {http://arxiv.org/abs/1312.5330} {arXiv:1312.5330 [hep-ph]} \BibitemShut
  {NoStop}%
%%CITATION = ARXIV:1312.5330;%%
\bibitem [{\citenamefont {Batra}\ and\ \citenamefont
  {Chacko}(2008)}]{Batra:2007iz}%
  \BibitemOpen
  \bibfield  {author} {\bibinfo {author} {\bibfnamefont {P.}~\bibnamefont
  {Batra}}\ and\ \bibinfo {author} {\bibfnamefont {Z.}~\bibnamefont {Chacko}},\
  }\href {\doibase 10.1103/PhysRevD.77.055015} {\bibfield  {journal} {\bibinfo
  {journal} {Phys.Rev.}\ }\textbf {\bibinfo {volume} {D77}},\ \bibinfo {pages}
  {055015} (\bibinfo {year} {2008})},\ \Eprint {http://arxiv.org/abs/0710.0333}
  {arXiv:0710.0333 [hep-ph]} \BibitemShut {NoStop}%
%%CITATION = ARXIV:0710.0333;%%
\bibitem [{\citenamefont {Wess}\ and\ \citenamefont
  {Zumino}(1971)}]{Wess:1971yu}%
  \BibitemOpen
  \bibfield  {author} {\bibinfo {author} {\bibfnamefont {J.}~\bibnamefont
  {Wess}}\ and\ \bibinfo {author} {\bibfnamefont {B.}~\bibnamefont {Zumino}},\
  }\href {\doibase 10.1016/0370-2693(71)90582-X} {\bibfield  {journal}
  {\bibinfo  {journal} {Phys.Lett.}\ }\textbf {\bibinfo {volume} {B37}},\
  \bibinfo {pages} {95} (\bibinfo {year} {1971})}\BibitemShut {NoStop}%
%%CITATION = PHLTA,B37,95;%%
\bibitem [{\citenamefont {Witten}(1984)}]{Witten:1983ar}%
  \BibitemOpen
  \bibfield  {author} {\bibinfo {author} {\bibfnamefont {E.}~\bibnamefont
  {Witten}},\ }\href {\doibase 10.1007/BF01215276} {\bibfield  {journal}
  {\bibinfo  {journal} {Commun.Math.Phys.}\ }\textbf {\bibinfo {volume} {92}},\
  \bibinfo {pages} {455} (\bibinfo {year} {1984})}\BibitemShut {NoStop}%
%%CITATION = CMPHA,92,455;%%
\bibitem [{\citenamefont {Witten}(1983)}]{Witten:1983tw}%
  \BibitemOpen
  \bibfield  {author} {\bibinfo {author} {\bibfnamefont {E.}~\bibnamefont
  {Witten}},\ }\href {\doibase 10.1016/0550-3213(83)90063-9} {\bibfield
  {journal} {\bibinfo  {journal} {Nucl.Phys.}\ }\textbf {\bibinfo {volume}
  {B223}},\ \bibinfo {pages} {422} (\bibinfo {year} {1983})}\BibitemShut
  {NoStop}%
%%CITATION = NUPHA,B223,422;%%
\bibitem [{\citenamefont {{CMS Collaboration}}(2014)}]{CMS:2014ega}%
  \BibitemOpen
  \bibfield  {author} {\bibinfo {author} {\bibnamefont {{CMS Collaboration}}},\
  }\href@noop {} {\bibfield  {journal} {\bibinfo  {journal}
  {CMS-PAS-HIG-14-009}\ } (\bibinfo {year} {2014})}\BibitemShut {NoStop}%
%%CITATION = CMS-PAS-HIG-14-009 ETC.;%%
\bibitem [{\citenamefont {Lee}\ and\ \citenamefont
  {Weinberg}(1977)}]{Lee:1977ua}%
  \BibitemOpen
  \bibfield  {author} {\bibinfo {author} {\bibfnamefont {B.~W.}\ \bibnamefont
  {Lee}}\ and\ \bibinfo {author} {\bibfnamefont {S.}~\bibnamefont {Weinberg}},\
  }\href {\doibase 10.1103/PhysRevLett.39.165} {\bibfield  {journal} {\bibinfo
  {journal} {Phys.Rev.Lett.}\ }\textbf {\bibinfo {volume} {39}},\ \bibinfo
  {pages} {165} (\bibinfo {year} {1977})}\BibitemShut {NoStop}%
%%CITATION = PRLTA,39,165;%%
\bibitem [{\citenamefont {Gondolo}\ and\ \citenamefont
  {Gelmini}(1991)}]{Gondolo:1990dk}%
  \BibitemOpen
  \bibfield  {author} {\bibinfo {author} {\bibfnamefont {P.}~\bibnamefont
  {Gondolo}}\ and\ \bibinfo {author} {\bibfnamefont {G.}~\bibnamefont
  {Gelmini}},\ }\href {\doibase 10.1016/0550-3213(91)90438-4} {\bibfield
  {journal} {\bibinfo  {journal} {Nucl.Phys.}\ }\textbf {\bibinfo {volume}
  {B360}},\ \bibinfo {pages} {145} (\bibinfo {year} {1991})}\BibitemShut
  {NoStop}%
%%CITATION = NUPHA,B360,145;%%
\bibitem [{\citenamefont {Ade}\ \emph {et~al.}(2014)\citenamefont {Ade} \emph
  {et~al.}}]{Ade:2013zuv}%
  \BibitemOpen
  \bibfield  {author} {\bibinfo {author} {\bibfnamefont {P.}~\bibnamefont
  {Ade}} \emph {et~al.} (\bibinfo {collaboration} {Planck Collaboration}),\
  }\href {\doibase 10.1051/0004-6361/201321591} {\bibfield  {journal} {\bibinfo
   {journal} {Astron.Astrophys.}\ } (\bibinfo {year} {2014}),\
  10.1051/0004-6361/201321591},\ \Eprint {http://arxiv.org/abs/1303.5076}
  {arXiv:1303.5076 [astro-ph.CO]} \BibitemShut {NoStop}%
%%CITATION = ARXIV:1303.5076;%%
\bibitem [{\citenamefont {Akerib}\ \emph {et~al.}(2014)\citenamefont {Akerib}
  \emph {et~al.}}]{Akerib:2013tjd}%
  \BibitemOpen
  \bibfield  {author} {\bibinfo {author} {\bibfnamefont {D.}~\bibnamefont
  {Akerib}} \emph {et~al.} (\bibinfo {collaboration} {LUX Collaboration}),\
  }\href {\doibase 10.1103/PhysRevLett.112.091303} {\bibfield  {journal}
  {\bibinfo  {journal} {Phys.Rev.Lett.}\ }\textbf {\bibinfo {volume} {112}},\
  \bibinfo {pages} {091303} (\bibinfo {year} {2014})},\ \Eprint
  {http://arxiv.org/abs/1310.8214} {arXiv:1310.8214 [astro-ph.CO]} \BibitemShut
  {NoStop}%
%%CITATION = ARXIV:1310.8214;%%
\bibitem [{\citenamefont {Cline}\ \emph {et~al.}(2013)\citenamefont {Cline},
  \citenamefont {Kainulainen}, \citenamefont {Scott},\ and\ \citenamefont
  {Weniger}}]{Cline:2013gha}%
  \BibitemOpen
  \bibfield  {author} {\bibinfo {author} {\bibfnamefont {J.~M.}\ \bibnamefont
  {Cline}}, \bibinfo {author} {\bibfnamefont {K.}~\bibnamefont {Kainulainen}},
  \bibinfo {author} {\bibfnamefont {P.}~\bibnamefont {Scott}}, \ and\ \bibinfo
  {author} {\bibfnamefont {C.}~\bibnamefont {Weniger}},\ }\href {\doibase
  10.1103/PhysRevD.88.055025} {\bibfield  {journal} {\bibinfo  {journal}
  {Phys.Rev.}\ }\textbf {\bibinfo {volume} {D88}},\ \bibinfo {pages} {055025}
  (\bibinfo {year} {2013})},\ \Eprint {http://arxiv.org/abs/1306.4710}
  {arXiv:1306.4710 [hep-ph]} \BibitemShut {NoStop}%
%%CITATION = ARXIV:1306.4710;%%
\bibitem [{\citenamefont {Alanne}\ \emph {et~al.}(2014)\citenamefont {Alanne},
  \citenamefont {Tuominen},\ and\ \citenamefont {Vaskonen}}]{Alanne:2014bra}%
  \BibitemOpen
  \bibfield  {author} {\bibinfo {author} {\bibfnamefont {T.}~\bibnamefont
  {Alanne}}, \bibinfo {author} {\bibfnamefont {K.}~\bibnamefont {Tuominen}}, \
  and\ \bibinfo {author} {\bibfnamefont {V.}~\bibnamefont {Vaskonen}},\ }\href
  {\doibase 10.1016/j.nuclphysb.2014.11.001} {\bibfield  {journal} {\bibinfo
  {journal} {Nucl.Phys.}\ }\textbf {\bibinfo {volume} {B889}},\ \bibinfo
  {pages} {692} (\bibinfo {year} {2014})},\ \Eprint
  {http://arxiv.org/abs/1407.0688} {arXiv:1407.0688 [hep-ph]} \BibitemShut
  {NoStop}%
%%CITATION = ARXIV:1407.0688;%%
\bibitem [{\citenamefont {Aprile}\ \emph {et~al.}(2011)\citenamefont {Aprile}
  \emph {et~al.}}]{Aprile:2011hi}%
  \BibitemOpen
  \bibfield  {author} {\bibinfo {author} {\bibfnamefont {E.}~\bibnamefont
  {Aprile}} \emph {et~al.} (\bibinfo {collaboration} {XENON100
  Collaboration}),\ }\href {\doibase 10.1103/PhysRevLett.107.131302} {\bibfield
   {journal} {\bibinfo  {journal} {Phys.Rev.Lett.}\ }\textbf {\bibinfo {volume}
  {107}},\ \bibinfo {pages} {131302} (\bibinfo {year} {2011})},\ \Eprint
  {http://arxiv.org/abs/1104.2549} {arXiv:1104.2549 [astro-ph.CO]} \BibitemShut
  {NoStop}%
%%CITATION = ARXIV:1104.2549;%%
\bibitem [{\citenamefont {Ackermann}\ \emph {et~al.}(2014)\citenamefont
  {Ackermann} \emph {et~al.}}]{Ackermann:2013yva}%
  \BibitemOpen
  \bibfield  {author} {\bibinfo {author} {\bibfnamefont {M.}~\bibnamefont
  {Ackermann}} \emph {et~al.} (\bibinfo {collaboration} {Fermi-LAT
  Collaboration}),\ }\href {\doibase 10.1103/PhysRevD.89.042001} {\bibfield
  {journal} {\bibinfo  {journal} {Phys.Rev.}\ }\textbf {\bibinfo {volume}
  {D89}},\ \bibinfo {pages} {042001} (\bibinfo {year} {2014})},\ \Eprint
  {http://arxiv.org/abs/1310.0828} {arXiv:1310.0828 [astro-ph.HE]} \BibitemShut
  {NoStop}%
%%CITATION = ARXIV:1310.0828;%%
\bibitem [{\citenamefont {Navarro}\ \emph {et~al.}(1997)\citenamefont
  {Navarro}, \citenamefont {Frenk},\ and\ \citenamefont
  {White}}]{Navarro:1996gj}%
  \BibitemOpen
  \bibfield  {author} {\bibinfo {author} {\bibfnamefont {J.~F.}\ \bibnamefont
  {Navarro}}, \bibinfo {author} {\bibfnamefont {C.~S.}\ \bibnamefont {Frenk}},
  \ and\ \bibinfo {author} {\bibfnamefont {S.~D.}\ \bibnamefont {White}},\
  }\href {\doibase 10.1086/304888} {\bibfield  {journal} {\bibinfo  {journal}
  {Astrophys.J.}\ }\textbf {\bibinfo {volume} {490}},\ \bibinfo {pages} {493}
  (\bibinfo {year} {1997})},\ \Eprint {http://arxiv.org/abs/astro-ph/9611107}
  {arXiv:astro-ph/9611107 [astro-ph]} \BibitemShut {NoStop}%
%%CITATION = ASTRO-PH/9611107;%%
\bibitem [{\citenamefont {Anderson}()}]{Anderson}%
  \BibitemOpen
  \bibfield  {author} {\bibinfo {author} {\bibfnamefont {B.}~\bibnamefont
  {Anderson}} (\bibinfo {collaboration} {Fermi-LAT Collaboration}),\
  }\href@noop {} {\enquote {\bibinfo {title} {A search for dark matter
  annihilation in dwarf spheroidal galaxies with pass 8 data},}\ }\bibinfo
  {note} {5th Fermi Symposium, October 24, 2014}\BibitemShut {NoStop}%
\bibitem [{\citenamefont {Litim}\ and\ \citenamefont
  {Sannino}(2014)}]{Litim:2014uca}%
  \BibitemOpen
  \bibfield  {author} {\bibinfo {author} {\bibfnamefont {D.~F.}\ \bibnamefont
  {Litim}}\ and\ \bibinfo {author} {\bibfnamefont {F.}~\bibnamefont
  {Sannino}},\ }\href@noop {} {\  (\bibinfo {year} {2014})},\ \Eprint
  {http://arxiv.org/abs/1406.2337} {arXiv:1406.2337 [hep-th]} \BibitemShut
  {NoStop}%
%%CITATION = ARXIV:1406.2337;%%
\end{thebibliography}%
%%%%%%%%%%%%%%%%%%%%%%%%%%%%%%%%%%

\end{document}